\documentclass[acmtog]{acmart}
\acmSubmissionID{561}

\usepackage{booktabs} % For formal tables

\usepackage{physics}
\usepackage{graphicx}
\usepackage{amsmath}
\usepackage{epstopdf}
\usepackage{siunitx}

\usepackage[ruled]{algorithm2e} % For algorithms

\SetAlFnt{\small}
\SetAlCapFnt{\small}
\SetAlCapNameFnt{\small}
\SetAlCapHSkip{0pt}

\acmJournal{TOG}
\newcommand{\mydraft}{false}

\newcommand{\mb}{\mathbf}
\newcommand{\xx}{\mb{x}}

\newcommand{\yy}{\mb{y}}
\newcommand{\bb}{\mb{b}}

\renewcommand{\qq}{\mb{q}}

\ifdef{\vv}{\renewcommand{\vv}{\mb{v}}}{\newcommand{\vv}{\mb{v}}}
\newcommand{\XX}{\mb{X}}

\newcommand{\FF}{\mb{F}}

\newcommand{\pp}{\mb{p}}
\renewcommand{\ss}{\mb{s}}
\newcommand{\rr}{\mb{r}}
\newcommand{\NN}{\mb{N}}
\newcommand{\II}{\mb{I}}
\newcommand{\ff}{\mb{f}}
\newcommand{\VV}{\mb{V}}
\newcommand{\TT}{\mb{T}}

\newcommand{\UU}{\mb{U}}

\newcommand{\GG}{\mb{G}}

\newcommand{\ii}{\mb{i}}
\newcommand{\jj}{\mb{j}}

\newcommand{\WW}{\mb{W}}

\newcommand{\ww}{\mb{w}}

\newcommand{\nn}{\mb{n}}

\renewcommand{\gg}{\mb{g}}
\renewcommand{\SS}{\mb{S}}
\renewcommand{\tt}{\mb{t}}
\renewcommand{\AA}{\mb{A}}

\newcommand{\sm}{\boldsymbol \sigma}
\newcommand{\Sm}{\boldsymbol \Sigma}
\newcommand{\fm}{\boldsymbol \phi}

\newcommand{\intot}{\int_{\Omega_t}}
\newcommand{\intst}{\int_{\partial\Omega_t}}

\begin{document}
% Title portion
\title{A Momentum-Conserving Implicit Material Point Method for Surface Energies with Spatial Gradients}
\begin{teaserfigure}
    \centering
    \includegraphics[draft=\mydraft,width=\textwidth,height=.25\textwidth]{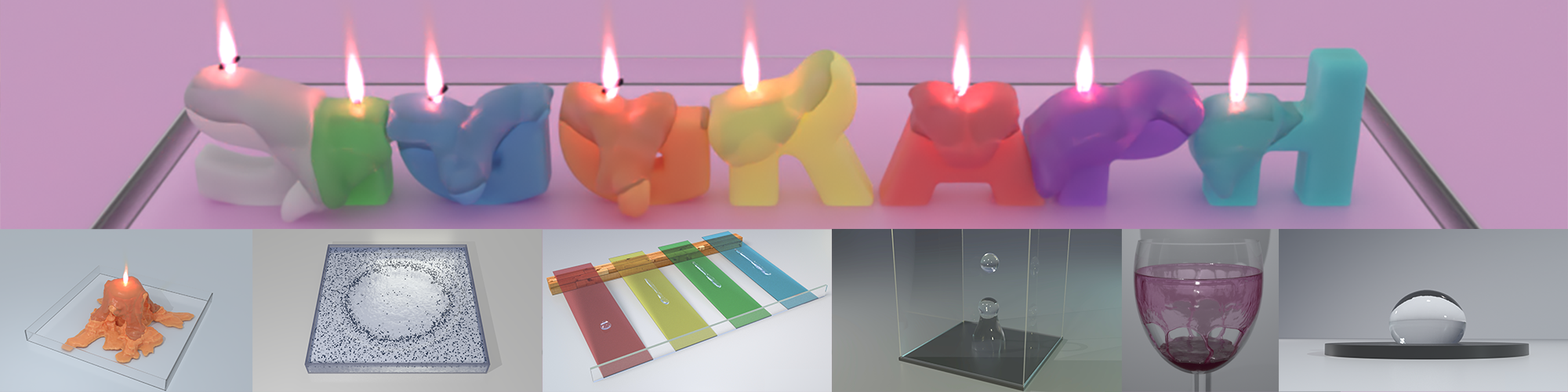} % teaser image is 3:2 aspect ratio e.g. 1500x1000px
    \caption{Our method enables the simulation of a wide variety of thermomechanical and surface-tension-driven effects. \textit{(Top)} Letter-shaped candles melt and interact.  \textit{(Bottom)} A large melting candle; soap spreading on a water surface; water droplets falling and streaking on ramps; partial rebound of a water droplet impact; wine settling in a glass; a water droplet settling on a hydrophobic surface.}
\end{teaserfigure}

% DO NOT ENTER AUTHOR INFORMATION FOR ANONYMOUS TECHNICAL PAPER SUBMISSIONS TO SIGGRAPH 2019!
\author{Jingyu Chen}
\affiliation{
	\institution{UCLA}
	\country{USA}
}

\author{Victoria Kala}
\affiliation{
	\institution{UCLA}
	\country{USA}
}

\author{Alan Marquez-Razon}
\affiliation{
	\institution{UCLA}
	\country{USA}
}

\author{Elias Gueidon}
\affiliation{
	\institution{UCLA}
	\country{USA}
}

\author{David A. B. Hyde}
\affiliation{
	\institution{UCLA}
	\country{USA}
}

\author{Joseph Teran}
\affiliation{
	\institution{UC Davis}
	\country{USA}
}

%\affiliation{%
%  \institution{Inria Paris-Rocquencourt}
%\email{beranger@inria.fr}
%\affiliation{%
% \institution{Rajiv Gandhi University}
%\email{aprna_patel@rguhs.ac.in}
%\affiliation{%
%  \institution{Tsinghua University}
%  \streetaddress{30 Shuangqing Rd}
%  \country{China}
%}
%\email{chan0345@tsinghua.edu.cn}
%\author{Ting Yan}
%\affiliation{%
%  \institution{Eaton Innovation Center}
%  \city{Prague}
%  \country{Czech Republic}}
%\email{yanting02@gmail.com}
%\author{Tian He}
%\affiliation{%
%  \institution{University of Virginia}
%  \department{School of Engineering}
%  \city{Charlottesville}
%  \state{VA}
%  \postcode{22903}
%  \country{USA}
%}
%\affiliation{%
%  \institution{University of Minnesota}
%\email{tinghe@uva.edu}
%\author{Chengdu Huang}
%\author{John A. Stankovic}
%\author{Tarek F. Abdelzaher}
%\affiliation{%
%  \institution{University of Virginia}
%  \department{School of Engineering}
%  \city{Charlottesville}
%  \state{VA}
%  \postcode{22903}
%  \country{USA}
%}

%\renewcommand\shortauthors{Zhou, G. et al}

\begin{abstract}
We present a novel Material Point Method (MPM) discretization of surface tension forces that arise from spatially varying surface energies. These variations typically arise from surface energy dependence on temperature and/or concentration. Furthermore, since the surface energy is an interfacial property depending on the types of materials on either side of an interface, spatial variation is required for modeling the contact angle at the triple junction between a liquid, solid and surrounding air. Our discretization is based on the surface energy itself, rather than on the associated traction condition most commonly used for discretization with particle methods. Our energy based approach automatically captures surface gradients without the explicit need to resolve them as in traction condition based approaches. We include an implicit discretization of thermomechanical material coupling with a novel particle-based enforcement of Robin boundary conditions associated with convective heating. Lastly, we design a particle resampling approach needed to achieve perfect conservation of linear and angular momentum with Affine-Particle-In-Cell (APIC) \cite{jiang:2015:apic}. We show that our approach enables implicit time stepping for complex behaviors like the Marangoni effect and hydrophobicity/hydrophilicity. We demonstrate the robustness and utility of our method by simulating materials that exhibit highly diverse degrees of surface tension and thermomechanical effects, such as water, wine and wax.
\end{abstract}





\maketitle

\section{Introduction}

Surface tension driven flows like those in milk crowns \cite{zheng:2015:new}, droplet coalescence \cite{thurey:2010:multiscale,wojtan:2010:surface_tracking,da:2016:surface,yang:2016:enriching,li:2020:kinetic} and bubble formation \cite{zhu:2014:codimensional,da:2015:bubble,huang:2020:chemomechanical} comprise some of the most visually compelling fluid motions.
Although these effects are most dominant at small scales, increasing demand for realism in computer graphics applications requires modern solvers capable of resolving them. Indeed surface tension effects have been well examined in the computer graphics and broader computational physics literature.
We design a novel approach for simulating surface tension driven phenomena that arise from spatial variations in cohesion and adhesion forces at the interface between two liquids. This is often called the Marangoni effect \cite{scriven:1960:marangoni,venerus:2015:tears} and perhaps the most famous example is the tears of wine phenomenon \cite{thomson:1855:marangoni}. Other notable examples of the Marangoni effect include repulsive flows induced by a soap droplet on a water surface as well as the dynamics of molten waxes and metals \cite{langbein:2002:capillary,farahi:2004:microfluidic}.\\
\\
The spatial variation in the surface forces can be characterized in terms of the potential energy $\Psi^s$ associated with surface tension:
\begin{align}\label{eq:stpe}
\Psi^s=\int_\Gamma k^\sigma(\xx)ds(\xx).
\end{align}
Here the surface tension coefficient $k^\sigma$ is proportionate to the relative cohesion and adhesion at the interface between the two fluids.
Typically this coefficient is constant across the multi-material interface $\Gamma$, however with the Marangoni effect the coefficient varies with $\xx\in\Gamma$.
These variations are typically driven by temperature or concentration gradients and give rise to many subtle, but important visual behaviors where the variation typically causes fluid to flow away from low surface energy regions towards high surface energy regions.
For example with tears of wine, the spatial variations in the surface energy arise from inhomogeneity in the mixture of alcohol and water caused by the comparatively rapid evaporation of alcohol and high surface tension of water.
\\
\\
There are many existing techniques in the computational physics literature that resolve spatial variations in the surface tension. For particle-based methods like Smoothed Particle Hydrodynamics (SPH) \cite{monaghan:1992:smoothed} and Particle-In-Cell (PIC) \cite{harlow:1965:mac}, most of these approaches are based on the Continuum Surface Force (CSF) model of Brackbill et al. \shortcite{brackbill:1992:continuum}. Marangoni effects have not been addressed in computer graphics, other than by Huang et al. \shortcite{huang:2020:chemomechanical} where it was examined for material flows in soap films. To capture the Marangoni effect, most approaches do not work with the potential energy $\Psi^s$ in Equation~\eqref{eq:stpe} but instead base their discretization on its first variation.
This variation results in the interfacial traction condition 
\begin{align}
\tt&=k^\sigma \kappa \nn + \nabla^S k^\sigma.\label{eq:st_cond}
\end{align}
Here $\tt$ is the force per unit area due to surface tension at the interface $\Gamma$, $\nabla^S$ is the surface gradient operator at the interface and $\kappa$ and $\nn$ are the interfacial mean curvature and normal, respectively. The original CSF technique of Brackbill et al. \shortcite{brackbill:1992:continuum} resolves the mean curvature term in Equation~\eqref{eq:st_cond}, but not the surface gradient term. Tong and Browne \shortcite{tong:2014:thermocapillary} show that the CSF approach can be modified to resolve the surface gradient. However, while this and other other existing approaches in the SPH and PIC literature are capable of resolving the spatial variation, none support implicit time stepping for the surface tension forces.\\
\\
We build on the work of Hyde et al. \shortcite{hyde:2020:tension} and show that efficient implicit time stepping with Marangoni effects is achievable with PIC. As in Hyde et al. \shortcite{hyde:2020:tension} we observe that similarities with hyperelasticity suggest that the Material Point Method (MPM) \cite{sulsky:1994:history-materials} is the appropriate version of PIC. We show that by building our discretization from the energy in Equation~\eqref{eq:stpe} rather than the more commonly adopted traction condition in Equation~\eqref{eq:st_cond}, we can naturally compute the first and second variations of the potential needed when setting up and solving the nonlinear systems of equations associated with fully implicit temporal discretization. Interestingly, by basing our discretization on the energy in Equation~\eqref{eq:stpe}, we also show that no special treatment is required for the interfacial spatial gradient operator $\nabla^s$ as was done in e.g. \cite{tong:2014:thermocapillary}. Furthermore, we show that our approach to discretizing the Marangoni forces can also be used to impose the contact angle at liquid/solid/air interfaces \cite{young:1805:cohesion}. We show that this naturally allows for simulation of droplet streaking effects.\\
\\
While our method is a generalization of the MPM technique in Hyde et al.\ \shortcite{hyde:2020:tension}, we also improve on its core functionality. The Hyde et al. \shortcite{hyde:2020:tension} approach is characterized by the introduction of additional surface tension particles at each time step which are designed to represent the liquid interface $\Gamma$ and its area weighted boundary normals. These surface tension particles are temporary and are deleted at the end of the time step to prevent excessive growth in particle count or macroscopic particle resampling. However, the surface tension particles are massless to prevent violation of conservation of particle mass and momentum. We show that this breaks perfect conservation of grid linear and angular momentum when particles introduce grid nodes with no mass that the surface tension forces will act upon. This is an infrequent occurrence, but breaks the otherwise perfect conservation of grid linear and angular momentum expected with conservative MPM forces. We design a novel mass and momentum resampling technique that, with the introduction of two new types of temporary particles, can restore perfect conservation of grid linear and angular momentum. We call these additional temporary particles balance particles. We show that our novel resampling provides improved behavior over the original approach of Hyde et al. \shortcite{hyde:2020:tension}, even in the case of standard, non-Marangoni surface tension effects. Furthermore, although other resampling techniques exists for PIC methods \cite{edwards:2012:hopic,yue:2015:mpm-plasticity,gao:2017:ampm}, we note that ours is the first to guarantee perfect conservation when using generalized particle velocities associated with the Affine Particle-in-Cell (APIC) method \cite{jiang:2016:course,jiang:2015:apic,fu:2017:poly}.\\
\\
Since variations in surface energy are typically based on temperature and/or concentration gradients, we couple our surface tension coefficients with thermodynamically driven quantities. Furthermore, we resolve solid to liquid and liquid to solid phase changes as a function of temperature since many Marangoni effects arise from melting and cooling. Notably, we show that our novel conservative resampling naturally improves discretization of Robin and Neumann boundary conditions on the interface $\Gamma$ needed for convection/diffusion of temperature and concentration. In summary, our primary contributions are:
\begin{itemize}
\item A novel implicit MPM discretization of spatially varying surface tension forces.
\item A momentum-conserving particle resampling technique for particles near the surface tension liquid interface.
\item An implicit MPM discretization of the convection/diffusion evolution of temperature/concentration coupled to the surface tension coefficient including a novel particle-based Robin boundary condition.
\end{itemize}

%%%%% move this figure to the remapping section
\begin{figure}
	\includegraphics[draft=\mydraft,width=\columnwidth,trim={10px 0 10px 0}]{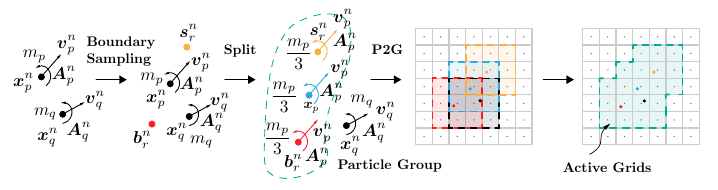}
	\caption{Splitting. After surface particles (yellow) are created, the mass and momentum of the interior MPM particles (blue) that are closest to the surface particles are immediately distributed. Particles in each particle group are assigned equal mass. 
	MPM particles (black) that are not paired with any surface particles remain intact for the splitting process. 
	Surface particles (yellow) and balance particles (red) are assigned the same linear velocity and affine velocity of the original particle (blue).}
	\label{fig:remapping_split}
\end{figure}

%%%%%

\section{Related Work}
We discuss relevant particle-based techniques for simulating Marangoni and surface tension effects, contact angle imposition, thermodynamic evolution of temperature and/or concentration, as well as resampling techniques in particle-based methods.

\paragraph{Particle Methods:}
Particle-based methods are very effective for computer graphics applications requiring discretization of surface tension forces. Hyde et al. \shortcite{hyde:2020:tension} provide a thorough discussion of the state of the art. Our approach utilizes the particle-based MPM \cite{sulsky:1994:history-materials,devaucorbeil:2020:review} PIC technique, largely due to its natural ability to handle self collision \cite{guo:2018:mpmshell,jiang:2017:cloth,fei:2018:mpmcloth,fei:2017:hair}, topology change \cite{wang:2019:fracture,wolper:2020:anisompm,wolper:2019:cdmpm}, diverse materials \cite{yue:2015:mpm-plasticity,stomakhin:2013:snow,ram:2015:mpm,daviet:2016:smp,wang:2020:parallel,klar:2016:des,schreck:2020:anisompm} as well as implicit time stepping with elasticity \cite{stomakhin:2013:snow,fei:2018:mpmcloth,wang:2020:hierarchical,fang:2019:silly}.
We additionally use the APIC method \cite{jiang:2016:course,jiang:2015:apic,fu:2017:poly} for its conservation properties and beneficial suppression of noise. Note that our mass and momentum remapping technique is designed to work in the context of the APIC techniques where particles store generalized velocity information.\\
\\
SPH is very effective for resolving Marangoni effects. The approaches of Tong and Browne \shortcite{tong:2014:thermocapillary} and Hopp-Hirschler et al.\ \shortcite{hopphirschler:2018:thermocapillary} are indicative of the state of the art. Most SPH works rely on the CSF surface tension model of Brackbill et al.\ \shortcite{brackbill:1992:continuum}, which transforms surface tension traction into a volumetric force that is only non-zero along (numerically smeared) material interfaces. CSF approaches generally derive surface normal and curvature estimates as gradients of color functions, which can be very sensitive to particle distribution. Also, CSF forces are not exactly conservative \cite{hyde:2020:tension}. SPH can also be used to simulate the convection and diffusion of temperature/concentration that gives rise to the spatial variation in surface energy in the Marangoni effect. Hu and Eberhard \shortcite{hu:2017:thermomechanically} simulate Marangoni convection in a melt pool during laser welding and Russell \shortcite{russell:2018:smoothed} does so in laser fusion additive manufacturing processes. Both approaches use SPH with Tong and Browne \shortcite{tong:2014:thermocapillary} for discretization of Equation~\eqref{eq:st_cond}. Although SPH is very effective for resolving Marangoni effects, all existing approaches utilize implicit treatment of Marangoni forces.

\begin{figure}
	\includegraphics[draft=\mydraft,width=\columnwidth,trim={10px 0 10px 0}]{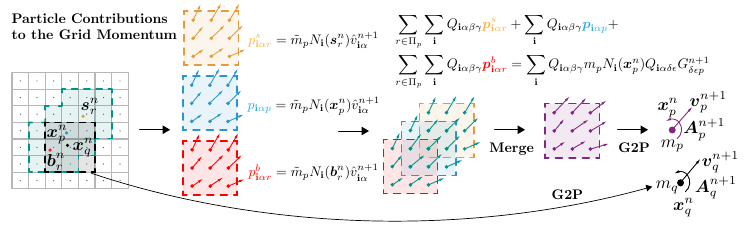}
	\caption{Merging. The merging process is a modified version of G2P. 
	For the particles that are not associated with surface particles (black), a regular G2P is performed.
	Among each particle group, we calculate each particle's contribution to the grid momentum and the generalized affine moments of their summed momenta about their center of mass.
	Then, we restore the mass of the original particle associated with the group prior to the split and compute its generalized affine inertia tensor from its grid mass distribution.
	Using the affine inertia tensor of the original particle, we compute generalized velocity of the particle after the merging from the generalized moments of the group.
	\label{fig:remapping_merge}
	}
\end{figure}

\paragraph{Marangoni effect and contact angle:}
The Marangoni effect is visually subtle and has not been resolved with particle-based methods in computer graphics applications. Perhaps the most visually compelling example of the Marangoni effect is the tears of wine phenomenon on the walls of a wine glass \cite{scriven:1960:marangoni,venerus:2015:tears}. Tears of wine were simulated in Azencot et al.\ \shortcite{azencot:2015:functional}, however the authors modeled the fluid using thin film equations under the lubrication approximation and did not model surface tension gradients. The fingering instabilities they observed are stated to occur due to the asymmetric nature of their initial conditions. The Marangoni effect was resolved by Huang et al.\ \shortcite{huang:2020:chemomechanical} recently with thin soap films to generate compelling dynamics of the characteristic rainbow patterns in bubbles. Relatedly, Ishida et al.\ \shortcite{ishida:2020:soap} simulated the evolution of soap films including effects of thin-film turbulence, draining, capillary waves, and evaporation. Outside of computer graphics, the Marangoni effect has been recently studied in works like Dukler et al.\ \shortcite{dukler:2020:theory}, which models undercompressive shocks in the Marangoni effect, and de Langavant et al.\ \shortcite{delangavant:2017:level}, which presents a spatially-adaptive level set approach for simulating surfactant-driven flows. Also, Nas and Tryggvason \shortcite{nas:2003:thermocapillary} simulated thermocapillary motion of bubbles and drops in flows with finite Marangoni numbers (flows with significant transport due to Marangoni effects).\\
\\
Our approach for the Marangoni effect also allows for imposition of contact angles at air/liquid/solid interfaces. This effect is important for visual realism when simulating droplets of water in contact with solid objects like the ground. Contact angles are influenced by the hydrophilicity/hydrophobicity of the surface on which these droplets move or rest \cite{cassie:1944:wettability,johnson:1964:contact}. Wang et al.\ \shortcite{wang:2007:solving} solve General Shallow Wave Equations, including surface tension boundary conditions and the virtual surface method of Wang et al.\ \shortcite{wang:2005:drops}, in order to model contact angles and hydrophilicity. Yang et al.\ \shortcite{yang:2016:versatile} use a pairwise force model \cite{tartakovsky:2005:modeling} for handling contact angles in their SPH treatment of fluid-fluid and solid-fluid interfaces. Clausen et al.\ \shortcite{clausen:2013:simulating} also consider the relation between surface tension and contact angles in their Lagrangian finite element approach.

\paragraph{Particle Resampling:}

Reseeding or resampling particles is a common concern in various simulation methods; generally speaking, particles need to be distributed with sufficient density near dynamic areas of flow or deformation in order to accurately resolve the dynamics of the system \cite{ando:2012:sheets,losasso:2008:twc,narain:2010:granular-materials}.
Edwards and Bridson \shortcite{edwards:2012:hopic} use a non-conservative random sampling scheme to seed and reseed particles with PIC.
Pauly et al.\ \shortcite{pauly:2005:fracture} resample to preserve detail near cracks/fracture, but their resampling does not attempt to conserve momentum. Yue et al.\ \shortcite{yue:2015:mpm-plasticity} use Poisson disk sampling to insert new points in low-density regions and merge points that are too close to one another with MPM. However, their resampling method is not demonstrated to be momentum-conserving. A conservative variant of this split-and-merge approach is applied in Gao et al.\ \shortcite{gao:2017:ampm} where mass and linear momentum are conserved during particle splitting and merging. However, angular momentum conservation is not conserved. Furthermore, these techniques use PIC not APIC and neither is designed to guarantee conservation with the generalized velocity state in APIC techniques \cite{jiang:2016:course,jiang:2015:apic,fu:2017:poly}.

\paragraph{Thermomechanical Effects:}

Thermodynamic effects in visual simulation date back to at least Terzopoulos et al.\ \shortcite{terzopoulos:1991:heating}.
More recently, melting and resolidification for objects like melting candles have been simulated using various methods, including Lattice Boltzmann \cite{wang:2012:multiple} and SPH \cite{paiva:2009:particle,lenaerts:2009:architecture}, though these results leave room for improved visual and physical fidelity.
FLIP methods have also been used for thermodynamic problems, such as Gao et al.\ \shortcite{gao:2017:heat}, which adapts the latent heat model from Stomakhin et al.\ \shortcite{stomakhin:2014:augmented-mpm}. 
Condensation and evaporation of water were considered in several works based on SPH \cite{hochstetter:2017:evaporation,zhang:2017:condensation}.
SPH was also applied to the problem of simulating boiling bubbles in Gu and Yang \shortcite{gu:2016:boiling}, which models heat conduction, convection and mass transfer.
Recently, particle-based thermodynamics models were incorporated into an SPH snow solver \cite{gissler:2020:snow}, with temperature-dependent material properties such as the Young's modulus.
In another vein, Pirk et al.\ \shortcite{pirk:2017:wood} used position-based dynamics and Cosserat physics to simulate combustion of tree branches, including models for moisture and charring.
Yang et al.\ \shortcite{yang:2017:unified} used the Cahn-Hilliard and Allen-Cahn equations to evolve a continuous phase variable for materials treated with their phase-field method.
Maeshima et al.\ \shortcite{maeshima:2020:particle} considered particle-scale explicit MPM modeling for additive manufacturing (selective laser sintering) that included a latent heat model for phase transition.
For a detailed review of thermodynamical effects in graphics, we refer the reader to Stomakhin et al.\ \shortcite{stomakhin:2014:augmented-mpm}.

\paragraph{Surface Tension:}

Many methods for simulating non-Marangoni surface tension effects have been developed for computer graphics applications. We refer the reader to Hyde et al.\ \shortcite{hyde:2020:tension} for a detailed survey. More recently, Chen et al.\ \shortcite{chen:2020:subgrid} incorporated sub-cell-accurate surface tension forces in an Eulerian fluid framework based on integrating the mean curvature flow of the liquid interface (following Sussman and Ohta \shortcite{sussman:2009:stable}).
With an eye towards resolving codimensional flow features, such as thin sheets and filaments, Wang et al.\ \shortcite{wang:2020:codimensional} and Zhu et al.\ \shortcite{zhu:2014:codimensional} simulated surface tension forces using moving-least-squares particles and simplicial complexes, respectively.
Most related to the present work, Hyde et al.\ \shortcite{hyde:2020:tension} proposed an implicit material point method for simulating liquids with large surface energy, such as liquid metals.
Their surface tension formulation follows \cite{adamson:1967:physical,brackbill:1992:continuum,buscaglia:2011:variational} and incorporates a potential energy associated with surface tension into the MPM framework.
Material boundaries are sampled using massless MPM particles.% (we note that Narin \shortcite{nairn:2003:mpm} considered massless auxiliary MPM particles for tracking crack propagation in fracture).
%However, the technique of Hyde et al.\ \shortcite{hyde:2020:tension} does not conserve linear or angular momentum, leading to undesirable results (as will be shown in this paper).
%Their model also assumes constant surface tension coefficients, which prevents simulating phenomena like the Marangoni effect.
%Readers are referred to Hyde et al.\ \shortcite{hyde:2020:tension} for a more thorough review of surface tension.

\section{Governing Equations}
We first define the governing equations for thermomechanically driven phase change of hyperelastic solids and liquids with variable surface energy. As in Hyde et al. \shortcite{hyde:2020:tension} we also cover the updated Lagrangian kinematics. Lastly, we provide the variational form of the governing equations for use in MPM discretization. We note that throughout the document Greek subscripts are assumed to run from $0,1,\hdots,d-1$ for the dimension $d=2,3$ of the problem. Repeated Greek subscripts imply summation, while sums are explicitly indicated for Latin subscripts. Also, Latin subscripts in bold are used for multi-indices.
\subsection{Kinematics}\label{sec:kin}
We adopt the continuum assumption \cite{gonzalez:2008:continuum} and updated Lagrangian kinematics \cite{belytschko:2013:nonlinear} used in Hyde et al. \shortcite{hyde:2020:tension}. At time $t$ we associate our material with subsets $\Omega^t\subset\mathbb{R}^d$, $d=2,3$. We use $\Omega^0$ to denote the initial configuration of material with $\XX\in\Omega^0$ used to denote particles of material at time $t=0$. A flow map $\fm:\Omega^0\times[0,T]\rightarrow\mathbb{R}^d$ defines the material motion of particles $\XX\in\Omega^0$ to their time $t$ locations $\xx\in\Omega^t$ as $\fm(\XX,t)=\xx$. The Lagrangian velocity is defined by differentiating the flow map in time $\VV(\XX,t)=\frac{\partial \fm}{\partial t}(\XX,t)$.
\subsubsection{Eulerian and Updated Lagrangian Representations}
The Lagrangian velocity can be difficult to work with in practice since real world observations of material are made in $\Omega^t$ not $\Omega^0$. The Eulerian velocity $\vv:\Omega^t\rightarrow\mathbb{R}^d$ is what we observe in practice. The Eulerian velocity is defined in terms of the inverse flow map $\fm^{-1}(\cdot,t):\Omega^t\rightarrow\Omega^0$ as $\vv(\xx,t)=\VV(\fm^{-1}(\xx,t),t)$ where $\fm^{-1}(\xx,t)=\XX$. In general, we can use the flow map and its inverse to pull back quantities defined over $\Omega^t$ and push forward quantities defined over $\Omega^0$, respectively. For example, given $G:\Omega^0\rightarrow\mathbb{R}$, its push forward $g:\Omega^t\rightarrow\mathbb{R}$ is defined as $g(\xx)=G(\fm^{-1}(\xx,t))$. This process is related to the material derivative operator $\frac{D}{Dt}$ where $\frac{Dg}{Dt}(\xx,t)=\frac{\partial G}{\partial t}(\fm^{-1}(\xx,t))=\frac{\partial g}{\partial t}(\xx,t) + \sum_{\alpha=0}^{d-1}\frac{\partial g}{\partial x_\alpha}(\xx,t)v_\alpha(\xx,t)$ (see e.g. \cite{gonzalez:2008:continuum} for more detail).\\
\\
In the updated Lagrangian formalism \cite{belytschko:2013:nonlinear} we write quantities over an intermediate configuration of material $\Omega^s$ with $0\leq s<t$. For example, we can define $\hat{g}(\cdot,s):\Omega^s\rightarrow\mathbb{R}$ as $\hat{g}(\tilde{\xx},s)=G(\fm^{-1}(\tilde{\xx},s))$ for $\tilde{\xx}\in\Omega^s$. As shown in Hyde et al. \shortcite{hyde:2020:tension}, this is particularly useful when discretizing momentum balance using its variational form. The key observation is that the updated Lagrangian velocity can be written as $\hat{\vv}(\tilde{\xx},s,t)=\VV(\fm^{-1}(\tilde{\xx},s),t)=\vv(\hat{\fm}(\tilde{\xx},s,t),t)$ with $\hat{\fm}(\tilde{\xx},s,t)=\fm(\fm^{-1}(\tilde{\xx},s),t))$ for $\tilde{\xx}\in\Omega^s$. Intuitively, $\hat{\fm}(\cdot,s,t):\Omega^s\rightarrow\Omega^t$ is the mapping from the time $s$ configuration to the time $t$ configuration induced by the flow map. This has a simple relation to the material derivative as $\frac{\partial \hat{\vv}}{\partial t}(\tilde{\xx},s,t)=\frac{\partial \VV}{\partial t}(\fm^{-1}(\tilde{\xx},s),t)=\frac{D\vv}{Dt}(\hat{\fm}(\tilde{\xx},s,t),t)$. As in Hyde et al. \shortcite{hyde:2020:tension} we will generally use upper case for Lagrangian quantities, lower case for Eulerian quantities and hat superscripts for updated Lagrangian quantities.

\subsubsection{Deformation Gradient}
The deformation gradient $\FF=\frac{\partial \fm}{\partial \XX}$ is defined by differentiating the flow map in space and can be used to quantify the amount of deformation local to a material point. We use $J=\det(\FF)$ to denote the deformation gradient determinant. $J$ represents the amount of volumetric dilation at a material point. Furthermore, it is used when changing variables with integration. We also make use of similar notation for the $\hat{\fm}$ mapping from $\Omega^s$ to $\Omega^t$, i.e. $\hat{\FF}=\frac{\partial \hat{\fm}}{\partial \tilde\xx}$, $\hat{J}=\det\big(\hat\FF\big)$. 

\subsection{Conservation of Mass and Momentum} Our governing equations primarily consist of conservation of mass and momentum which can be expressed as
\begin{align}\label{eq:balance}
\rho\frac{D\vv}{Dt}&=\nabla\cdot\sm + \rho \gg, \ \frac{D\rho}{Dt}=-\rho\nabla\cdot\vv, \ \xx\in\Omega^t
\end{align}
where $\rho$ is the Eulerian mass density, $\vv$ is the Eulerian material velocity, $\sm$ is the Cauchy stress and $\gg$ is gravitational acceleration. Boundary conditions for these equations are associated with a free surface for solid material, surface tension for liquids and/or prescribed velocity conditions. We use $\partial\Omega^t_N$ to denote the portion of the time $t$ boundary subject to free surface or surface tension conditions and $\partial\Omega^t_D$ to denote the portion of the boundary with Dirichlet velocity boundary conditions. Free surface conditions and surface tension boundary conditions are expressed as
\begin{align}
\sm\nn=\tt, \xx \in \partial \Omega^t_N \label{eq:tbc}
\end{align}
where $\tt=\mb{0}$ for free surface conditions and $\tt = k^\sigma \kappa\nn + \nabla^S k^\sigma$ from Equation~\eqref{eq:st_cond} for surface tension conditions. Velocity boundary conditions may be written as
\begin{align}\label{eq:vbc}
\vv \cdot \nn = v_{\text{bc}}^n, \xx \in \partial \Omega_D^t.
\end{align}
\subsubsection{Constitutive Models}
Each material point is either a solid or liquid depending on the thermomechanical evolution. For liquids, the Cauchy stress $\sm$ is defined in terms of pressure and viscous stress:
\begin{align*}
\sm=-p\II + \mu\left(\frac{\partial \vv}{\partial \xx} + \frac{\partial \vv}{\partial \xx}^T\right), \ p=-\frac{\partial \Psi^p}{\partial J},
\end{align*}
with $\Psi^p(J) = \frac{\lambda^l}{2}(J-1)^2$. Here $\lambda^l$ is the bulk modulus of the liquid and $\mu^l$ is its viscosity. For solids, the Cauchy stress is defined in terms of a hyperelastic potential energy density $\Psi^s$ as
\begin{align*}
\sm=\frac{1}{j}\frac{\partial \psi^h}{\partial \FF}\ff^T
\end{align*}
where $\ff(\xx,t)=\FF(\fm^{-1}(\xx,t),t)$ and $j(\xx,t)=J(\fm^{-1}(\xx,t),t)$ are the Eulerian deformation gradient and its determinant, respectively. We use the fixed-corotated constitutive model from \cite{stomakhin:2012:invertible} for $\psi^h$. This model is defined in terms of the polar SVD \cite{ITF04} of the deformation gradient $\FF=\UU\Sm\VV^T$ with
\begin{equation*}
\psi^h(\FF) = \mu^h \sum_{\alpha=0}^{d-1} (\sigma_\alpha - 1)^2 + \frac{\lambda^h}{2} (J-1)^2 ,
\end{equation*}
where the $\sigma_\alpha$ are the diagonal entries of $\Sm$ and $\mu^h,\lambda^h$ are the hyperelastic Lam\'e coefficients.

\subsection{Conservation of energy} We assume the internal energy of our materials consists of potential energy associated with surface tension, liquid pressure and hyperelasticity and thermal energy associated with material temperature. Conservation of energy together with thermodynamic considerations requires convection/diffusion of the material temperature \cite{gonzalez:2008:continuum} subject to Robin boundary conditions associated with convective heating by ambient material:
\begin{equation}
  \begin{aligned}
    \rho c_p \frac{DT}{Dt} &= K \Delta T + H \\
    K \grad T \cdot \nn &= - h (T - \bar{T}) + b .
    \label{eqn:heat}
  \end{aligned}
\end{equation}
Here $c_p$ is specific heat capacity, $T$ is temperature, $K$ is thermal diffusivity, $H$ is a source function, $\bar{T}$ is the temperature of ambient material and $\nn$ is the surface boundary normal. $h$ controls the rate of convective heating to the ambient temperature and $b$ represents the rate of boundary heating independent of the ambient material temperature $\bar{T}$.\\
\\
The total potential energy $\Psi$ in our material is as in Hyde et al. \shortcite{hyde:2020:tension}, however we include the spatial variation of the surface energy density in Equation~\eqref{eq:stpe} and the hyperelastic potential for solid regions:
\begin{align*}
\Psi(\fm(\cdot,t))=\Psi^{\sigma}(\fm(\cdot,t))+\Psi^{l}(\fm(\cdot,t))+\Psi^h(\fm(\cdot,t))+ \Psi^g(\fm(\cdot,t)).
\end{align*}
Here $\Psi^{\sigma}$ is the potential from surface tension, $\Psi^{l}$ is the potential from liquid pressure, $\Psi^h$ is the potential from solid hyperelasticity and $\Psi^g$ is the potential from gravity. As in typical MPM discretizations, our approach is designed in terms of these energies:
\begin{align*}
\Psi^g(\fm(\cdot,t)&=\int_{\Omega^0}R\gg\cdot\fm J d\XX, \ \Psi^{\sigma}(\fm(\cdot,t))=\int_{\partial \Omega^t} k^{\sigma}(\xx,t) ds(\xx)\\\
\Psi^{l}(\fm(\cdot,t))&=\int_{\Omega^0}\frac{\lambda^l}{2}\left(J-1\right)^2d\XX, \ \Psi^{h}(\fm(\cdot,t))=\int_{\Omega^0}\psi^h(\FF)d\XX.
\end{align*}
Here $R$ is the pull back (see Section~\ref{sec:kin}) of the mass density $\rho$. Note that in the expression for the surface tension potential it is useful to change variables using the updated Lagrangian view as in Hyde et al. \shortcite{hyde:2020:tension}:
{\small
\begin{align}
\Psi^{\sigma}(\fm(\cdot,t))&=\int_{\partial \Omega^t} k^{\sigma}(\xx,t) ds(\xx)=\int_{\partial \Omega^s} k^{\sigma}(\hat{\fm}(\tilde{\xx},s,t),t) |\hat{J}\hat{\FF}^{-T}\tilde{\nn}| ds(\tilde{\xx}).\label{eq:pestul}
\end{align}}
Here $\tilde{\nn}$ is the outward unit normal at a point on the boundary of $\Omega^s$ and the expression $ds(\xx)=|\hat{J}\hat{\FF}^{-T}\tilde{\nn}|ds(\tilde{\xx})$ arises by a change of variables from an integral over $\Omega^t$ to one over $\Omega^s$.  Notably, the spatial variation in $k^{\sigma}$ does not require a major modification of the Hyde et al. \shortcite{hyde:2020:tension} approach.

\subsection{Variational Form of Momentum Balance}
The strong form of momentum balance in Equation~\eqref{eq:balance}, together with the traction (Equation~\eqref{eq:tbc}) and Dirichlet velocity boundary conditions (Equation~\eqref{eq:vbc}), is equivalent to a variational form that is useful when discretizing our governing equations using MPM. To derive the variational form, we take the dot product of Equation~\eqref{eq:balance} with an arbitrary function $\ww:\Omega^t\rightarrow\mathbb{R}^d$ satisfying $\ww\cdot\nn=0$ for $\xx\in\partial\Omega^t_D$ and integrate over the domain $\Omega^t$, applying integration by parts where appropriate. Requiring that the Dirichlet velocity conditions in Equation~\eqref{eq:tbc} hold together with the following integral equations for all functions $\ww$ is equivalent to the strong form, assuming sufficient solution regularity:
\begin{align}\label{eq:varmom}
\int_{\Omega^t}\rho\frac{Dv_\alpha}{Dt}w_\alpha d\xx=-\frac{d}{d\epsilon}PE(0;\ww) - \mu^l \int_{
\Omega^t}\epsilon^v_{\alpha\beta}\epsilon^w_{\alpha\beta}d\xx.
\end{align}
Here $\epsilon^w=\frac{1}{2}\left(\frac{\partial w_\alpha}{\partial x_\beta} + \frac{\partial w_\beta}{\partial x_\alpha}\right)$ and $\epsilon^v=\frac{1}{2}\left(\frac{\partial v_\alpha}{\partial x_\beta} + \frac{\partial v_\beta}{\partial x_\alpha}\right)$ and
\begin{align*}
\textrm{PE}(\epsilon;\ww)=\Psi(\fm(\cdot,t)+\epsilon\WW).
\end{align*}
Here $\WW$ is the pull back of $\ww$ (see Section~\ref{sec:kin}). Note that this notation is rather subtle for the surface tension potential energy. For clarification,
\begin{align*}
\Psi^{\sigma}(\fm(\cdot,t)+\epsilon \WW)&=\int_{\partial \Omega^s} k^{\sigma}(\hat{\fm}+\epsilon\hat{\ww}) |\hat{J}_{\epsilon,
\hat{\ww}}\hat{\FF}_{^{\epsilon,
\hat{\ww}}}^{-T}\tilde{\nn}| ds(\tilde{\xx}),
\end{align*}
where $\hat{\ww}(\tilde{\xx},s,t)=\ww(\hat{\fm}(\tilde{\xx},s,t))$ is the updated Lagrangian pull back of $\ww$, $\hat{\FF}_{^{\epsilon,
\hat{\ww}}}=\frac{\partial \hat{\fm}+\epsilon\hat{\ww}}{\partial \tilde{\xx}}$ is the deformation gradient of the mapping $\hat{\fm}+\epsilon\hat{\ww}$ and $\hat{J}_{\epsilon,
\hat{\ww}}=\det\big( \hat{\FF}_{^{\epsilon,
\hat{\ww}}}\big)$ is its determinant. Lastly, for discretization purposes, in practice we change variables in the viscosity term
\begin{align}\label{eq:visc_ul}
\mu^l \int_{
\Omega^t}\epsilon^v_{\alpha\beta}\epsilon^w_{\alpha\beta}d\xx = \mu^l \int_{
\Omega^s}\hat{\epsilon}^v_{\alpha\beta}\hat{\epsilon}^w_{\alpha\beta} \hat{J} d\tilde{\xx}.
\end{align}\\
\\
We can similarly derive a variational form of the temperature evolution in Equation~\eqref{eqn:heat} by requiring 
\begin{equation}
	\label{eq:energy_weak}
	\begin{aligned}
	\intot \rho c_p \frac{DT}{Dt} q d\xx = &- \intst q h T  d\SS(\xx) + \intst q \left(h \bar{T}+b\right) d\SS(\xx) \\ 
		&+ \intot H q d\xx  - \intot \grad{q} \cdot K \grad{T} d\xx
	\end{aligned}
\end{equation}
for all functions $q:\Omega^t\rightarrow\mathbb{R}$.

\subsection{Thermomechanical Material Dependence and Phase Change}
The thermomechanical material dependence is modeled by allowing the surface tension coefficient $k^\sigma$, the liquid bulk modulus $\lambda^l$, the liquid viscosity $\mu^l$ and the hyperelastic Lam\'e coefficients $\mu^h,\lambda^h$ to vary with temperature. When $T$ exceeds a user-specified melting point $T_\text{melt}$, the solid phase is changed to liquid and the deformation gradient determinant $J$ is set to $1$.
Similarly, if liquid temperature drops below $T_\text{melt}$, The phase is updated to be hyperelastic solid and the deformation gradient $\FF$ is set to the identity matrix. In practice, resetting the deformation gradient and its determinant helps to prevent nonphysical popping when the material changes from solid to liquid and vice versa. We remark that incorporating a more sophisticated phase change model, such as a latent heat buffer, is potentially useful in future work \cite{stomakhin:2014:augmented-mpm}.

\subsection{Contact Angle} \label{sec:contact_angle}
The contact angle between a liquid, a solid boundary and the ambient air is governed by the Young equation \cite{young:1805:cohesion}. This expression relates the resting angle $\theta$ (measured through the liquid) of a liquid in contact with a solid surface to the surface tension coefficients between the liquid, solid and air phases:
\begin{equation}
k^\sigma_{SG} = k^\sigma_{SL} + k^\sigma_{LG} \cos(\theta).
\label{eqn:young}
\end{equation}
The surface tension coefficients are between the solid and gas phases, solid and liquid phases, and liquid and gas phases, respectively.
As in Clausen et al. \shortcite{clausen:2013:simulating}, we assume $k^\sigma_{SG}$ is negligible since we are using a free surface assumption and do not explicitly model the air. Under this assumption, the solid-liquid contact angle is determined by the surface tension ratio $-k^\sigma_{SL}/k^\sigma_{LG}$. We note that, while one would expect surface tension coefficients/energies to be positive, this ratio can be negative under the assumption of zero solid-gas surface tension. Furthermore, we note that utilizing this expression requires piecewise constant surface tension coefficients where the variation along the liquid boundary is based on which portion is in contact with the air and which is in contact with the solid. 
The distinct surface tension coefficients on different interfaces provide controllability of the spreading behavior of the liquid on the solid surface.

\section{Discretization}

As in Hyde et al.\ \shortcite{hyde:2020:tension}, we use MPM \cite{sulsky:1994:history-materials} and APIC \cite{jiang:2015:apic} to discretize the governing equations. The domain $\Omega^{t^n}$ at time $t^n$ is sampled using material points $\xx_p^n$.
These points also store approximations of the deformation gradient determinant $J_p^n$, constant velocity $\vv_p^n$, affine velocity $\AA_p^n$, volume $V_p^0$, mass $m_p=\rho(\xx_p^0,t^0)V_p^0$, temperature $T_p^n$, and temperature gradient $\grad T_p^n$. We also make use of a uniform background grid with spacing $\Delta x$ when discretizing momentum updates. To advance our state to time $t^{n+1}$, we use the following steps:
\begin{enumerate}
\item Resample particle boundary for surface tension and Robin boundary temperature conditions.
\item P2G: Conservative transfer of momentum and temperature from particles to grid.
\item Update of grid momentum and temperature.
\item G2P: Conservative transfer of momentum and temperature from grid to particles.
\end{enumerate}
\subsection{Conservative Surface Particle Resampling}

\begin{figure}[ht]
    \centering
    \includegraphics[draft=\mydraft,width=\columnwidth]{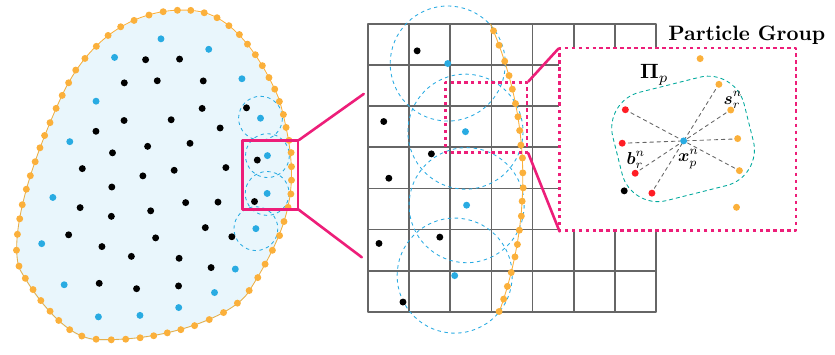}
    \caption{A portion of an MPM fluid in the simulation domain.
    Surface particles (yellow) are sampled on faces of the zero isocontour of the level set formed by unioning spherical level sets around each MPM particle.
    Each surface particle generates an associated balance particle (red) such that the closest MPM particle (blue) to a boundary particle lies on the midpoint of a line segment between the surface particle and balance particle.
    A single blue particle at $\boldsymbol{x}_p$ may be paired with multiple surface particles and balance particles, and they are considered to be in a particle group $\boldsymbol{\Pi}_p$.
    MPM particles that are not associated with any surface tension particles are marked as black.}
    \label{fig:balance-diagram}
\end{figure}

The integrals associated with the surface tension energy in Equation~\eqref{eq:pestul} and the Robin temperature condition in Equation~\eqref{eq:energy_weak} are done over the boundary of the domain. We follow Hyde et al. \shortcite{hyde:2020:tension} and introduce special particles to cover the boundary in order to serve as quadrature points for these integrals. As in Hyde et al. \shortcite{hyde:2020:tension} these particles are temporary and are removed at the end of the time step. However, while Hyde et al.\ \shortcite{hyde:2020:tension} used massless surface particles, we design a novel conservative mass and momentum resampling for surface particles. Massless particles easily allow for momentum conserving transfers from particle to grid and vice versa; however, they can lead to loss of conservation in the grid momentum update step. This occurs when there is a grid cell containing only massless particles. In this case, there are grid nodes with no mass that receive surface tension forces. These force components are then effectively thrown out since only grid nodes with mass will affect the end of time step particle momentum state (see Section \ref{sec:conservation-example}).\\
\\
We resolve this issue by assigning mass to each of the surface particles. However, to conserve total mass, some mass must be subtracted from interior MPM particles. Furthermore, changing the mass of existing particles also changes their momentum, which may lead to violation of conservation. In order to conserve mass, linear momentum and angular momentum, we introduce a new particle for each surface particle. We call these balance particles, and like surface particles they are temporary and will be removed at the end of the time step. We show that the introduction of these balance particles naturally allows for conservation both when they are created at the beginning of the time step and when they are removed at the end of the time step.
\subsubsection{Surface Particle Sampling}
\label{sec:boundary-quadrature}
\begin{figure}[ht]
    \centering
    \includegraphics[draft=\mydraft,width=.8\columnwidth,trim={120px 0 100px 0},clip]{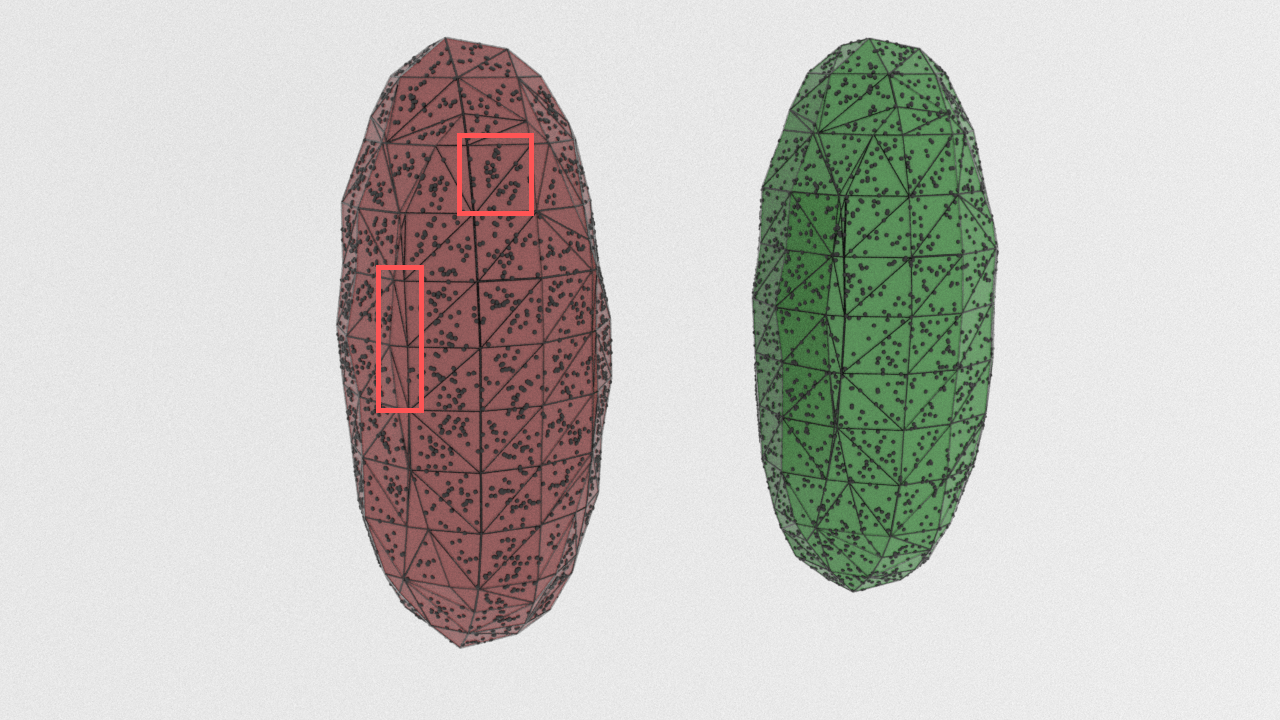}
    \caption{Isocontour and sampled boundary particles for an ellipsoid. \textit{(Left)} Using the method of Hyde et al.\ \shortcite{hyde:2020:tension}.  Note how low-quality triangles are undersampled and how sample points often clump near triangle centers.  \textit{(Right)} The present method, which does not suffer from similar issues.}
    \label{fig:sampling-comparison}
\end{figure}

We first introduce surface particles using the approach in Hyde et al.\ \shortcite{hyde:2020:tension}. A level set enclosing the interior MPM particles is defined as the union of spherical level sets defined around each interior MPM particle. Unlike Hyde et al.\ \shortcite{hyde:2020:tension}, we do not smooth or shift the unioned level set. We compute the zero isocontour of the level set using marching cubes \cite{chernyaev:1995:marching} and randomly sample surface particles along this explicit representation. In Hyde et al.\ \shortcite{hyde:2020:tension}, three-dimensional boundaries were sampled using a number of sample points proportional to the surface area of each triangle. Sample points were computed using uniform random barycentric weights, which leads to a non-uniform distribution of points in each triangle. Instead, we employ a strategy of per-triangle Monte Carlo sampling using a robust Poisson distribution, as described in Corsini et al.\ \shortcite{corsini:2012:sampling}, where uniform triangle sample points are generated according to Osada et al.\ \shortcite{osada:2002:shape}. We found that this gave better coverage of the boundary without generating particles that are too close together  (see Figure \ref{fig:sampling-comparison}). We note that radii for the particle level sets are taken to be $0.73\Delta x$ (slightly larger than $\frac{\sqrt{2}}{2\Delta x}$) in 2D and $0.867\Delta x$ (slightly larger than $\frac{\sqrt{3}}{2\Delta x}$) in 3D. This guarantees that even a single particle in isolation will always generate a level set zero isocontour that intersects the grid and will therefore always generate boundary sample points. Note also that as in Hyde et al. \shortcite{hyde:2020:tension}, we use the explicit marching cubes mesh of the zero isocontour to easily and accurately generate samples of area weighted normals $d\AA_r$ where $\sum |d\AA_r|\approx\int_\Omega^{t^n}d\xx$ are chosen with direction from the triangle normal and magnitude based on the number of samples in a given triangle and the triangle area.

%\begin{figure}[ht]
%    \includegraphics[draft=\mydraft,width=.32\columnwidth]{example-image-a}
%    \includegraphics[draft=\mydraft,width=.32\columnwidth]{example-image-b}
%    \includegraphics[draft=\mydraft,width=.32\columnwidth]{example-image-c}
%    \caption{A dropping ellipsoid with the method of \cite{hyde:2020:tension} (left), with our new remapping scheme (middle), and with our remapping scheme plus improved boundary sampling (right).  Both of our contributions appear to yield more symmetric, plausible results.}
%    \label{fig:compare-remapping-sampling}
%\end{figure}

\subsubsection{Balance Particle Sampling}
For each surface particle $\ss_r^n$, we additionally generate a balance particle $\bb_r^n$. First, we compute the closest interior MPM particle for each surface particle $\xx_{p(\ss_r^n)}^n$. Then we introduce the corresponding balance particle as
\begin{align}\label{eq:pbal}
\bb_r^n = \ss_r^n + 2\left(\xx_{p(\ss_r^n)}^n-\ss_r^n\right).
\end{align}
\subsubsection{Mass and Momentum Splitting}
After introducing the surface $\ss^n_r$ and balance $\bb^n_r$ particles, we assign them mass and momentum (see Figure~\ref{fig:remapping_split}). To achieve this in a conservative manner, we first partition the surface particles into particle groups $\Pi_p$ defined as the set of surface particle indices $r$ such that $\xx_p^n$ is the closest interior MPM particle to $\ss^n_r$ (see Figure~\eqref{fig:balance-diagram}). We assign the mass $m_p$ of the particle $\xx_p^n$ to the collection of $\xx_p^n$, $\ss^n_r$ and $\bb^n_r$ for $r\in\Pi_p$ uniformly by defining a mass of $\tilde{m}_p=\frac{m_p}{2|\Pi_p|+1}$ to each surface and balance point as well as to $\xx_p^n$. Here $|\Pi_p|$ is the number of elements in the set. This operation is effectively a split of the original particle $\xx_p^n$ with mass $m_p$ into a new collection of particles $\xx_p^n, \ss^n_r, \bb^n_r, \ r\in\Pi_p$ with masses $\tilde{m}_p$. This split trivially conserves the mass. Importantly, by construction of the balance particles (Equation~\eqref{eq:pbal}) we ensure that the center of mass of the collection is equal to the original particle $\xx_p^n$:
\begin{align}\label{eq:cons_mom}
\frac{1}{m_p} \left( \tilde{m}_p\xx_p^n+\sum_{r\in\Pi_p} \tilde{m}_p\ss^n_r + \tilde{m}_p\bb^n_r\right)=\xx_p^n.
\end{align}
With this particle distribution, conservation of linear and angular momentum can be achieved by simply assigning each new particle in the collection the velocity $\vv_p^n$ and affine velocity $\AA_p^n$ of the original particle $\xx_p^n$. We note that the conservation of the center of mass (Equation~\eqref{eq:cons_mom}) is essential for this simple constant velocity split to conserve linear and angular momentum (see \cite{anon:2021:tech_doc}).

\subsection{Transfer: P2G}
After the addition of the surface and balance particles, we transfer mass and momentum to the grid in the standard APIC \cite{jiang:2015:apic} way using their conservatively remapped mass and velocity state
\begin{align*}
m^n_{\ii}&=\sum_{p}\tilde{m}_p\left(N_{\ii}(\xx^n_p) + \sum_{r\in\Pi_p}N_{\ii}(\ss^n_r) + N_{\ii}(\bb^n_r) \right),\\
\quad m^n_{\ii}\vv^n_{\ii}&=\sum_{p}\tilde{m}_pN_{\ii}(\xx^n_p)\left(\vv^n_p+\AA^n_p(\xx_{\ii}-\xx^n_p)\right)\\
&+\sum_{p}\tilde{m}_p\sum_{r\in\Pi_p}N_{\ii}(\ss^n_r)\left(\vv^n_p+\AA^n_p(\xx_{\ii}-\ss^n_r)\right)\\
&+\sum_{p}\tilde{m}_p\sum_{r\in\Pi_p}N_{\ii}(\bb^n_r)\left(\vv^n_p+\AA^n_p(\xx_{\ii}-\bb^n_r)\right).
\end{align*}
Here $\NN_{\ii}(\xx) = \NN(\xx - \xx_{\ii})$ are quadratic B-splines defined over the uniform grid with $\xx_\ii$ living at cell centers \cite{stomakhin:2013:snow}. Note that for interior MPM particles far enough from the boundary that $\Pi_p=\varnothing$. This reduces to the standard APIC \cite{jiang:2015:apic} splat since $\tilde{m}_p=m_p^n$. We also transfer temperature from particles to grid using
\begin{align*}
T^n_{\ii\alpha} = \sum_{p}m_p N_{\ii}(\xx^n_p)(T_p^n + (x_{\ii\alpha}-x^n_{p\alpha})\nabla T^n_{p\alpha}).
\end{align*}
Note that for the temperature transfer, we only use surface particles to properly apply the thermal boundary conditions, and we do not use these particles to transfer mass-weighted temperature to the grid.

\subsection{Grid Momentum and Temperature Update}
We discretize the governing equations in the standard MPM manner by using the particles as quadrature points in the variational forms. The interior MPM particles $\xx_p^n$ are used for volume integrals and the surface particles $\ss_r^n$ are used for surface integrals. By choosing $s=t^n$, $t=t^{n+1}$ and by using grid discretized versions of $\hat{\ww}(\tilde{\xx})=\sum_\jj\ww_\jj N_\jj(\tilde{\xx})$, $\hat{\vv}(\tilde{\xx},t^n,t^{n+1})=\sum_\ii\hat{\vv}^{n+1}_\ii N_\ii(\tilde{\xx})$, $\hat{q}(\tilde{\xx})=\sum_\jj q_\jj N_\jj(\tilde{\xx})$ and $\hat{T}(\tilde{\xx})=\sum_\ii \hat{T}_\ii N_\ii(\tilde{\xx})$.
\subsubsection{Momentum Update}
As in Hyde et al.\ \shortcite{hyde:2020:tension}, the grid momentum update is derived from Equation~\eqref{eq:varmom}:
\begin{align}
m^n_\ii\frac{\hat{\vv}^{n+1}_\ii-\vv^n_\ii}{\Delta t}&=\ff_\ii(\xx+\Delta t\hat{\qq})+m^n_\ii\gg,\\
\ff_{\ii}(\hat \xx) &= -\frac{\partial e}{\partial \hat \xx_{\ii}}(\hat{\xx}) -\mu^l\sum_p \boldsymbol\epsilon^v(\hat{\xx};\xx_p^n)\left(\frac{\partial N_\ii}{\partial \xx} \label{eq:disc_force}(\xx_p^n)\right)^T V_p^n,
\end{align}
where $\ff_{\ii}$ is the force on grid node $\ii$ from potential energy and viscosity, $\epsilon^v(\hat{\xx};\xx_p^n)=\frac{1}{2}\left(\sum_\jj \hat{\xx}_\jj\frac{\partial N_\jj}{\partial \xx}(\xx_p^n) + \left(\hat{\xx}_\jj\frac{\partial N_\jj}{\partial \xx}(\xx_p^n)\right)^T\right)$ is the strain rate at $\xx_p^n$, $\gg$ is gravity, and $\hat \qq$ is either $0$ (for explicit time integration) or $\hat \vv^{n+1}$ (for backward Euler time integration).
$\xx$ represents the vector of all unmoved grid node positions $\xx_{\ii}$.
We use $e(\yy)$ to denote the discrete potential energy $\Psi$ where MPM and surface particles are used as quadrature points:
\begin{align*}
e(\yy)&=\sum_p \left(\psi^h(\FF_p(\hat{\yy})) + \frac{\lambda^l}{2}(J_p(\hat{\yy})-1)^2\right)V_p^0\\
&+ \sum_r k^\sigma(\ss_r^n)|\hat{J}_r(\hat{\yy})\hat{\FF}_r^{-T}(\hat{\yy})d\AA^n_r|,
\end{align*}
where, as in \cite{stomakhin:2013:snow}, $\FF_p(\hat{\yy})=\sum_\ii \yy_\ii \frac{\partial N_\ii}{\partial \xx}(\xx^n_p)\FF_p^n$ and as in Hyde et al. \shortcite{hyde:2020:tension}, $J_p(\hat{\yy})=\left(1-d+y_\alpha \frac{\partial N_\ii}{\partial x_\alpha}(\xx_p^n)\right)J_p^n$ and $\hat{\FF}_p(\yy)=\sum_\ii \yy_\ii\frac{\partial N_\ii}{\partial \xx}(\xx_p^n)$. With these conventions, the $\alpha$ component of the energy-based force on grid node $\ii$ is of the form
\begin{equation}
	\small{
	\begin{aligned}
	-\frac{\partial e}{\partial x_{\ii\alpha}}(\yy)&=-\sum_p \frac{\partial \psi^h}{\partial F_{\alpha\delta}}(\FF_p(\hat{\yy}))F^n_{p\gamma\delta}\frac{\partial N_\ii}{\partial x_{\gamma}}(\xx_p^n)V_p^0\\
	&-\sum_p\lambda^l(J_p(\yy)-1)\frac{\partial N_\ii}{\partial x_{\alpha}}(\xx_p^n)J_p^nV_p^0\\
	&-\sum_r k^\sigma(\ss_r^n)\frac{\partial |\det(\hat{\GG}_r)\hat{\GG}^{-T}_r d\AA^n_r|}{\partial \hat{G}_{\alpha\delta}}(\hat{\FF}_r(\hat{\yy}))\frac{\partial N_\ii}{\partial x_{\delta}}(\xx_p^n).
	\end{aligned}}\label{eq:fe}
\end{equation}
We note that the viscous contribution to the force in Equation~\eqref{eq:disc_force} is the same as in Ram et al. \shortcite{ram:2015:mpm}. We would expect $V_p^{n+1}$ in this term when deriving from Equation~\eqref{eq:visc_ul}, however we approximate it as $V_p^{n}$. This is advantageous since it makes the term linear; and since $\hat{J}_p(\hat{\xx})\approx 1$ from the liquid pressure and hyperelastic stress, it is not a poor approximation. Lastly, we note that the surface tension coefficient $k^\sigma(\ss_r^n)$ will typically get its spatial dependence from composition with a function of temperature $k^\sigma(\ss_r^n)=\tilde{k}^\sigma(\hat{T}(\ss_r^n))=\tilde{k}^\sigma(\TT^{s,n}_r)$.\\
\\
In the case of implicit time stepping with backward Euler ($\hat{\qq}=\hat{\vv}^{n+1}$), we use Netwon's method to solve the nonlinear systems of equations. This requires linearization of the grid forces associated with potential energy in Equation~\eqref{eq:fe}. We refer the reader to Stomakhin et al.\ \shortcite{stomakhin:2013:snow} and Hyde et al.\ \shortcite{hyde:2020:tension} for the expressions for these terms, as well as the definiteness fix used for surface tension contributions.

\subsubsection{Temperature Update}
We discretize Equation \ref{eq:energy_weak} in a similar manner which results in the following equations for the grid temperatures $T_\ii$:
\begin{align*}
	c_p m_{\ii} \frac{\hat{T}_{\ii}^{n+1} - T_{\ii}^n}{\Delta t} =
		&- \sum_p K \frac{\partial N_\ii}{\partial x_\alpha}(\xx_p^n) {\hat{T}_{\jj}^{n+1}} \frac{\partial N_\jj}{\partial x_\alpha}(\xx_p^n) V_p^n \\
		&- \sum_r h N_{\ii}(\ss_r^n) \hat{T}^{n+1}_{\jj} N_{\jj}(\ss_r^n) |d\AA_r^n| \\
		&+ \sum_r N_{\ii}(\ss_r^n) \left[ h \bar{T}(\ss_r^n) + b(\ss_r^n) \right] |d\AA_r^n| \\
		&+ \sum_p h N_{\ii}(\xx_p^n) H(\xx_p^n) V_p^n.
\end{align*}
Note that by using the surface particles $\ss_r^n$ as quadrature points in the variational form, the Robin boundary condition can be discretized naturally with minimal modification to the Laplacian and time derivative terms. Also note that inclusion of this term modifies both the matrix and the right side in the linear system for $\hat{T}_{\ii}^{n+1}$.
We found that performing constant extrapolation of interior particle temperatures to the surface particles provided better initial guesses for the linear solver.

\section{Transfer: G2P}
Once grid momentum and temperature have been updated, we transfer velocity and temperature back to the particles. For interior MPM particles with no associated surface or balance particles ($\Pi_p=\varnothing$), we transfer velocity, affine velocity and temperature from grid to particles in the standard APIC \cite{jiang:2015:apic} way:
\begin{equation*}
\vv^{n+1}_p=\sum_{\ii}N_\ii(\xx^n_p)\hat{\vv}^{n+1}_\ii,\quad \AA^{n+1}_p=\frac{4}{\Delta x^2}\sum_{\ii}N_\ii(\xx^n_p)\hat{\vv}^{n+1}_\ii(\xx_\ii-\xx^n_p)^T.
\end{equation*}
For interior MPM particles that were split with a collection of surface and balance particles ($\Pi_p\neq\varnothing$), more care must be taken since surface and balance particles will be deleted at the end of the time step. First, the particle is reassigned its initial mass $m_p$. Then we compute the portion of the grid momentum associated with each surface and balance particle associated with $p$ as
\begin{align*}
\pp^s_{\ii r}=\tilde{m}_pN_\ii(\ss^n_r)\hat{\vv}^{n+1}_\ii, \ \pp^b_{\ii r}=\tilde{m}_pN_\ii(\bb^n_r)\hat{\vv}^{n+1}_\ii, \ r\in\Pi_p.
\end{align*}
We then sum this with the split particle's share of the grid momentum to define the merged particle's share of the grid momentum
\begin{align*}
\pp_{\ii p}=\tilde{m}_pN_\ii(\xx^n_p)\hat{\vv}^{n+1}_\ii + \sum_{r\in\Pi_p} \pp^s_{\ii r} + \pp^b_{\ii r}.
\end{align*}
Note that the $\pp_{\ii p}$ may be nonzero for more grid nodes than the particle would normally splat to (see Figure~\ref{fig:remapping_merge}). We define the particle velocity from the total momentum by dividing by the mass $\vv_p^{n+1}=\frac{1}{m_p}\sum_\ii \pp_{\ii p}$. To define the affine particle velocity, we use a generalization of Fu et al. \shortcite{fu:2017:poly} and first compute the generalized affine moments $t_{p\beta\gamma}=\sum_\ii Q_{\vb{i} \alpha \beta \gamma} p_{\ii p\alpha}$ of the momentum distribution $p_{\ii p\alpha}$ where $Q_{\vb{i} \alpha \beta \gamma}=r_{\ii p\gamma} \delta_{\alpha\beta}$ is the $\alpha$ component of the $\beta \gamma$ linear mode at grid node $\ii$. Here $\rr_{\ii p}=\xx_\ii-\xx_p^n$ is the displacement from the center of mass of the distribution to the grid node $\xx_\ii$. We note that these moments are the generalizations of angular momentum to affine motion, as was observed in \cite{jiang:2015:apic}, however in our case we compute the moments from a potentially wider distribution of momenta $p_{\ii p\alpha}$. Lastly, to conserve angular momenta (see  \cite{anon:2021:tech_doc} for details), we define the affine velocity by inverting the generalized affine inertia tensor $\sum_\ii Q_{\vb{i} \alpha \gamma \delta} m_pN_\ii(\xx_p^n) Q_{\vb{i} \alpha \epsilon \tau}$ of the point $\xx_p^n$ using its merged mass distribution $m_p N_\ii(\xx_p^n)$. However, as noted in \cite{jiang:2015:apic}, the generalized inertia tensor $\frac{m_p{\Delta x}^2}{4}\II$ is constant diagonal when using quadratic B-splines for $N_\ii(\xx_p^n)$ and therefore the final affine velocity is $\AA_p^{n+1}=\frac{4}{m_p{\Delta x}^2}\tt_p$.\\
\\
Temperature and temperature gradients are transferred in the same way whether or not a MPM particle was split or not:
\begin{align*}
T_p^{n+1}=\sum_\ii \hat{T}^{n+1}_\ii N(\xx_\ii), \ \nabla T_p^{n+1}=\sum_\ii \hat{T}^{n+1}_\ii\nabla N(\xx_\ii).
\end{align*}

\section{Examples}

\subsection{Conservation}
\label{sec:conservation-example}

\begin{figure}[ht]
    \includegraphics[draft=\mydraft,width=0.49\columnwidth]{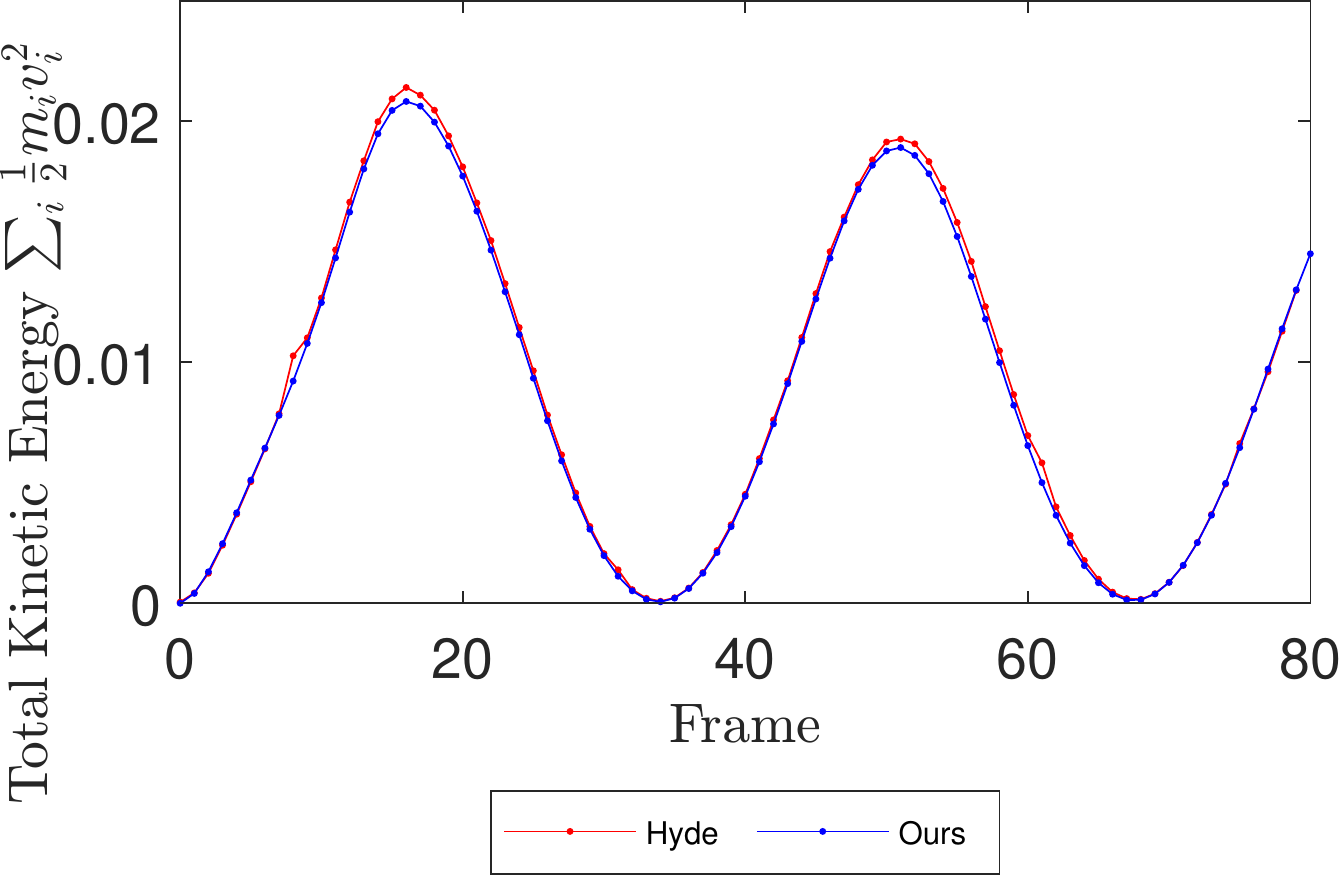}
    \includegraphics[draft=\mydraft,width=0.49\columnwidth]{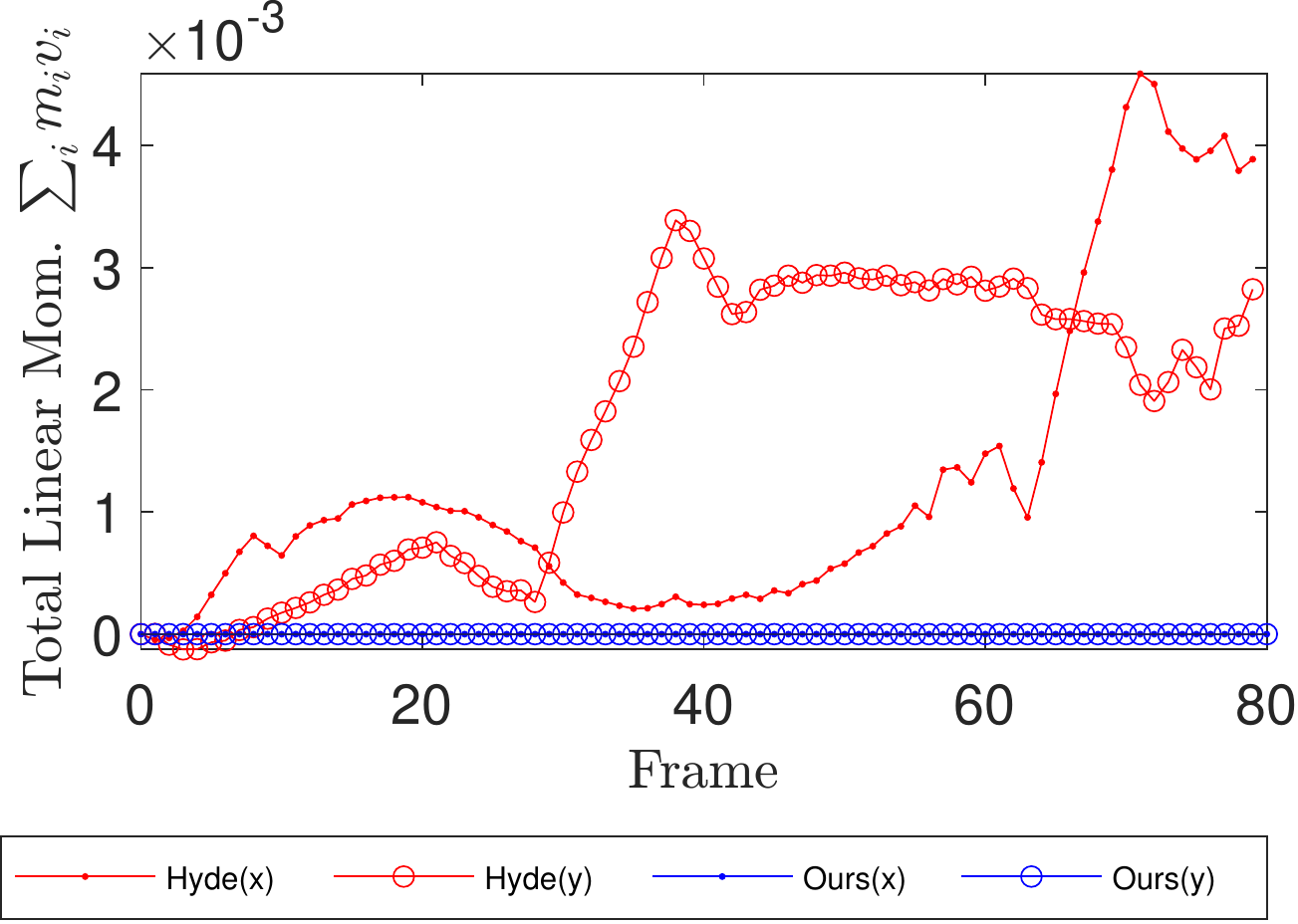}
    \includegraphics[draft=\mydraft,width=0.49\columnwidth]{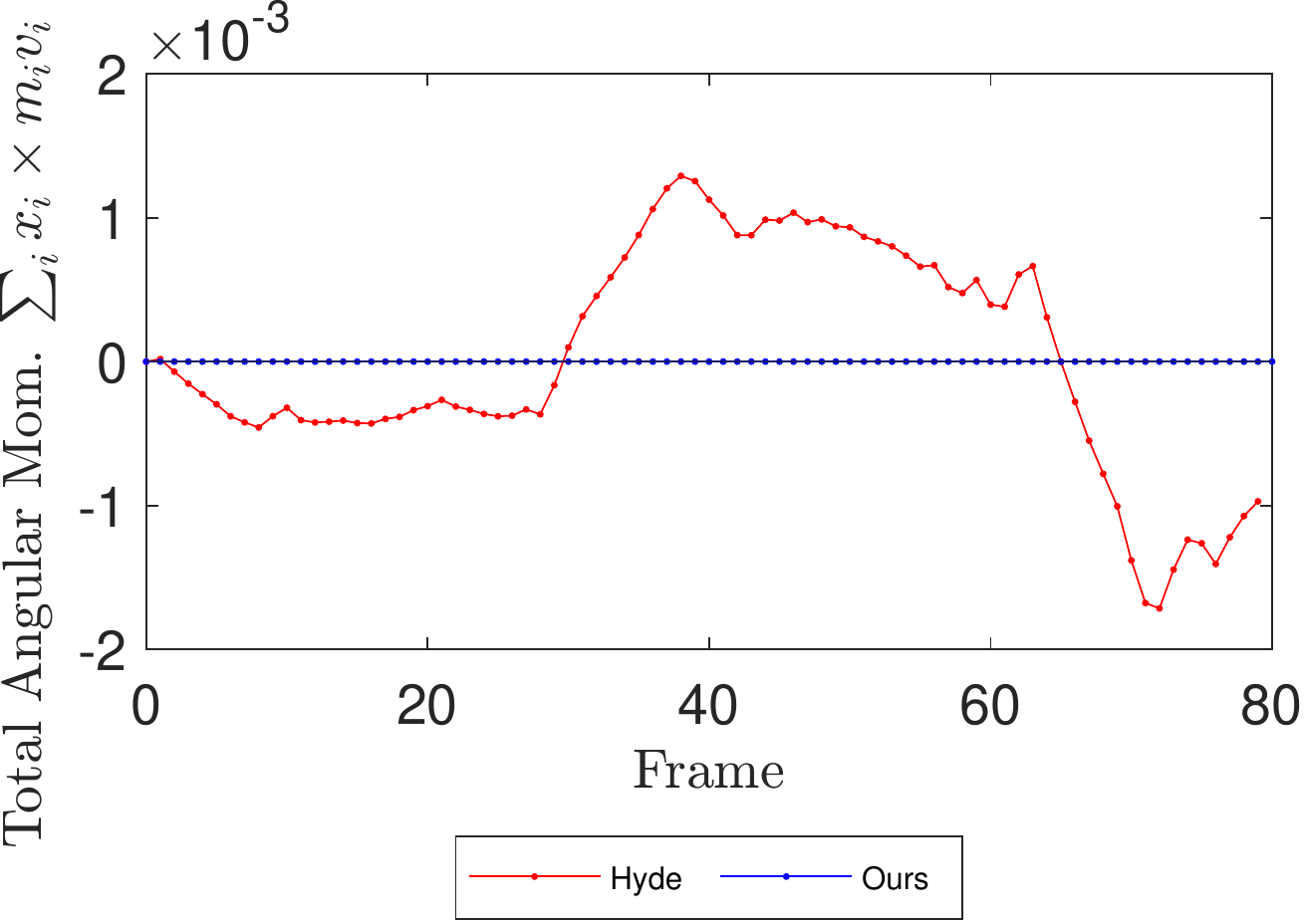}
    \includegraphics[draft=\mydraft,width=0.49\columnwidth]{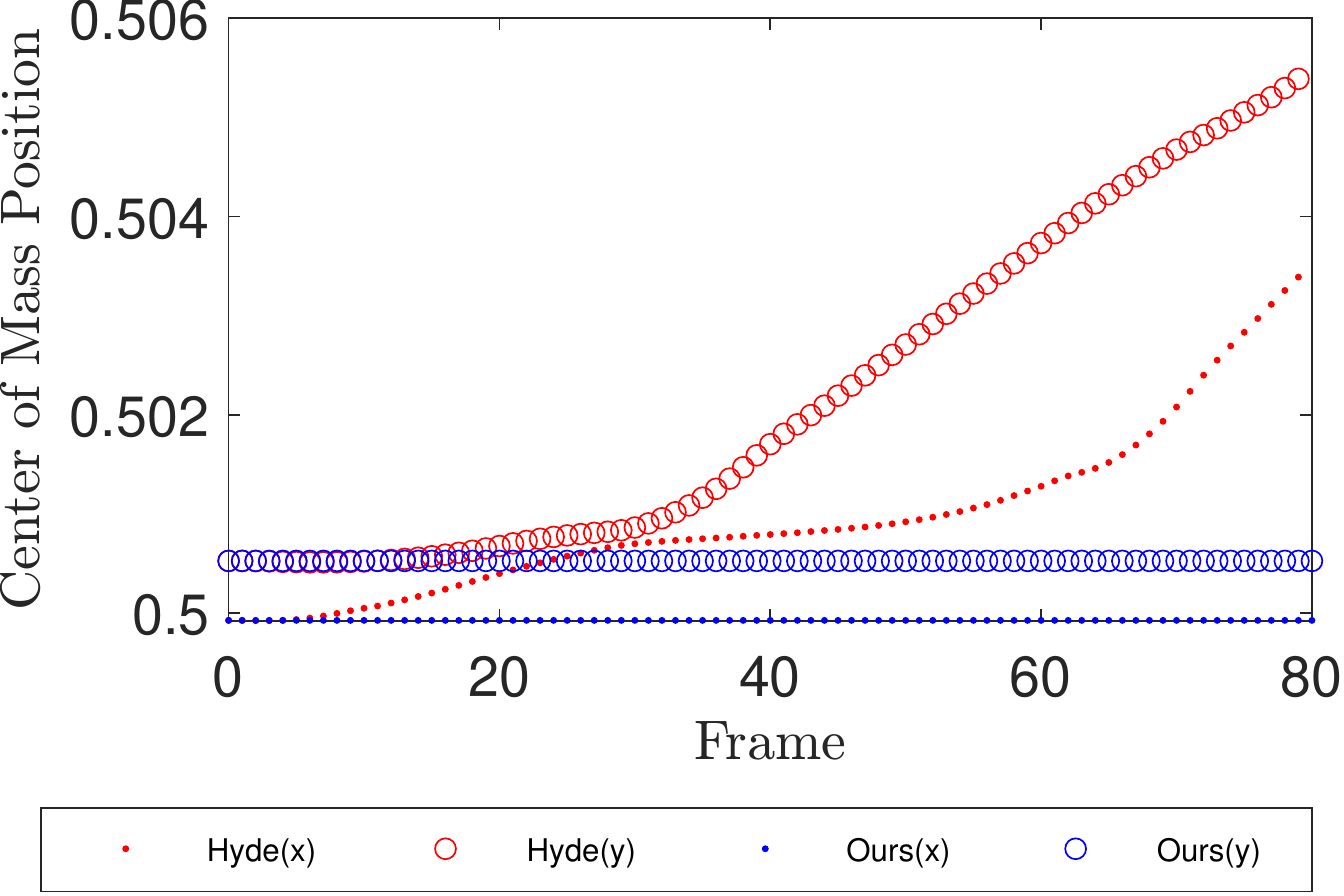}
    \caption{The present method (blue) conserves total mass, total linear and angular momentum, and center of mass, unlike Hyde et al.\ \shortcite{hyde:2020:tension} (red).  Both methods are energy-dissipative.}
    \label{fig:ellipse_conservation}
\end{figure}

To demonstrate our method's ability to fully conserve momentum and center of mass, we simulate a two-dimensional ellipse that oscillates under zero gravity due to surface tension forces, and we compare these results to those obtained using the method of Hyde et al.\ \shortcite{hyde:2020:tension}.
As seen in Figure \ref{fig:ellipse_conservation}, both methods are dissipative in terms of kinetic energy due to the APIC transfer \cite{jiang:2015:apic}.
However, the present technique perfectly conserves total linear momentum, total angular momentum, and the center of mass of the ellipse, unlike Hyde et al.\ \shortcite{hyde:2020:tension}.
In this example, explicit time stepping was used, and the CFL time step was restricted to be between \num{1e-4} and \num{1e-6} seconds. A grid resolution of $\Delta x = 1/63$ was used, along with a bulk modulus of $4166.67$ and a surface tension coefficient of $k^\sigma = 0.1$.
8 particles per cell were randomly sampled, then filtered to select only the ones inside the ellipse; this resulted in a center of mass for the ellipse at approximately $(0.4993, 0.5005)$.

\subsection{Droplet Impact on Dry Surface}

\begin{figure}[ht]
	\includegraphics[draft=\mydraft,width=0.95\columnwidth,trim={25px 100px 0  0},clip]{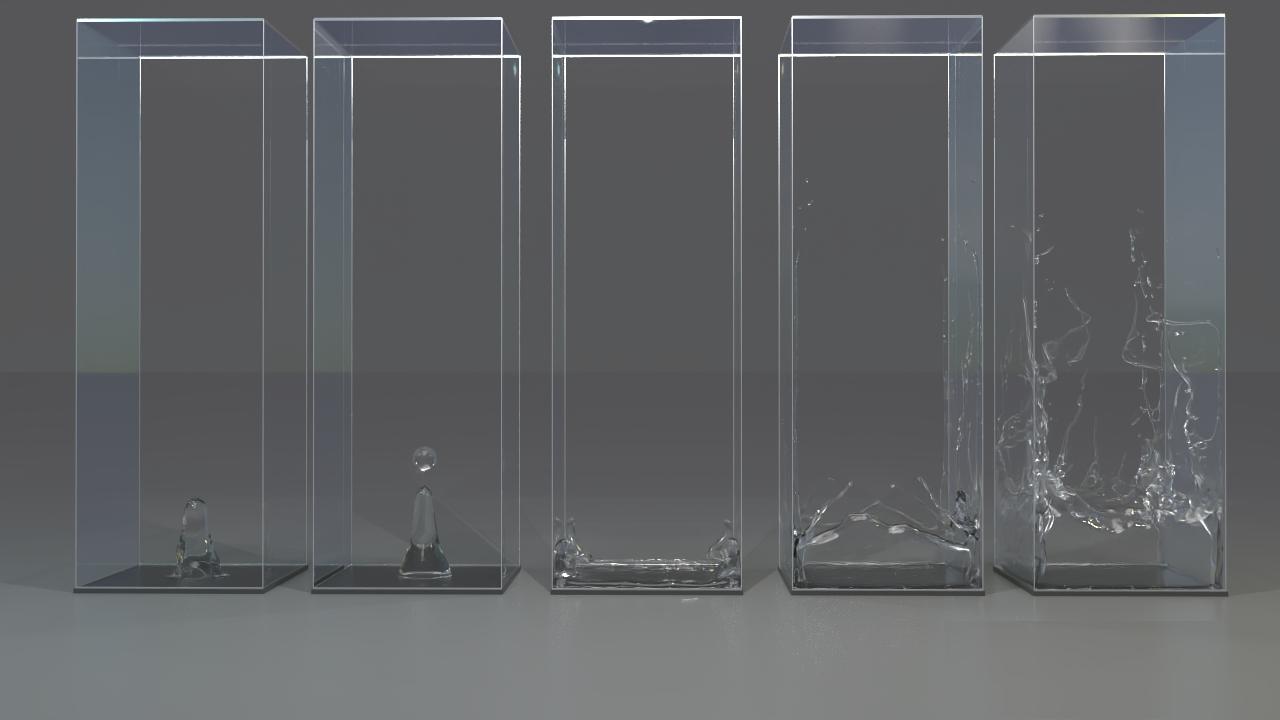}
	\includegraphics[draft=\mydraft,width=0.31\columnwidth,trim={300px 0 300px  0},clip]{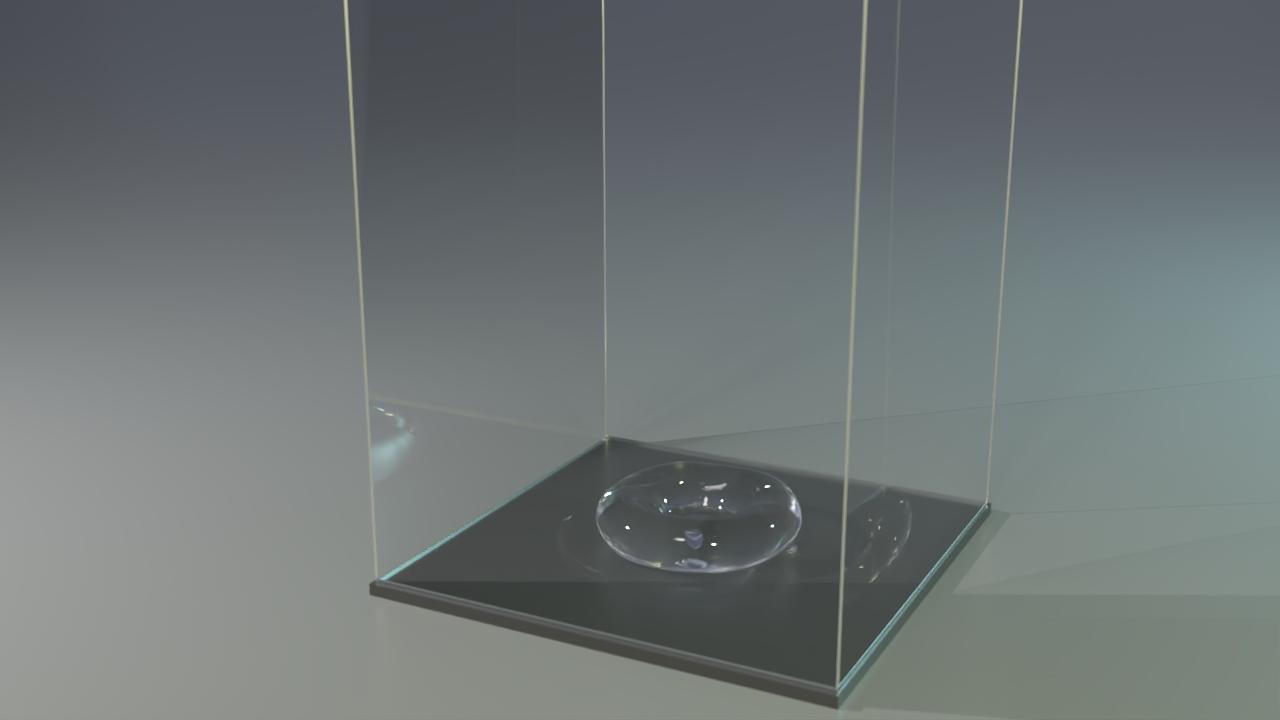}
	\includegraphics[draft=\mydraft,width=0.31\columnwidth,trim={300px 0 300px  0},clip]{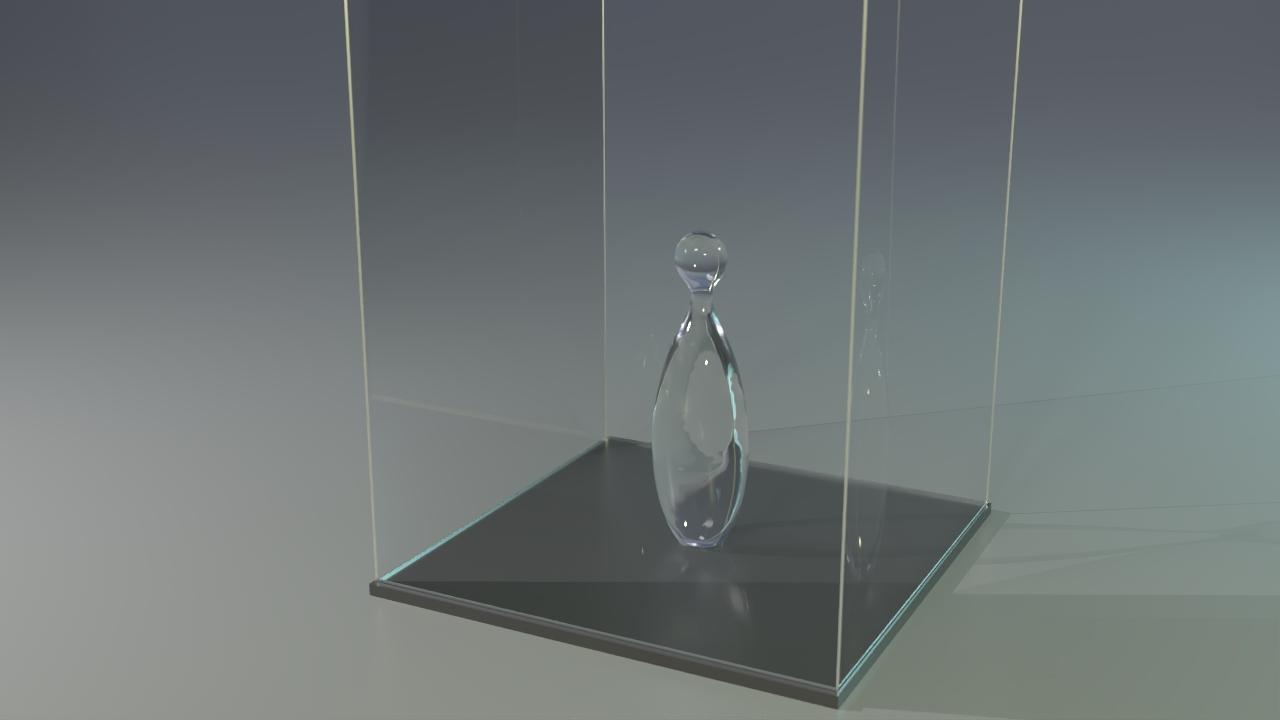}
	\includegraphics[draft=\mydraft,width=0.31\columnwidth,trim={300px 0 300px  0},clip]{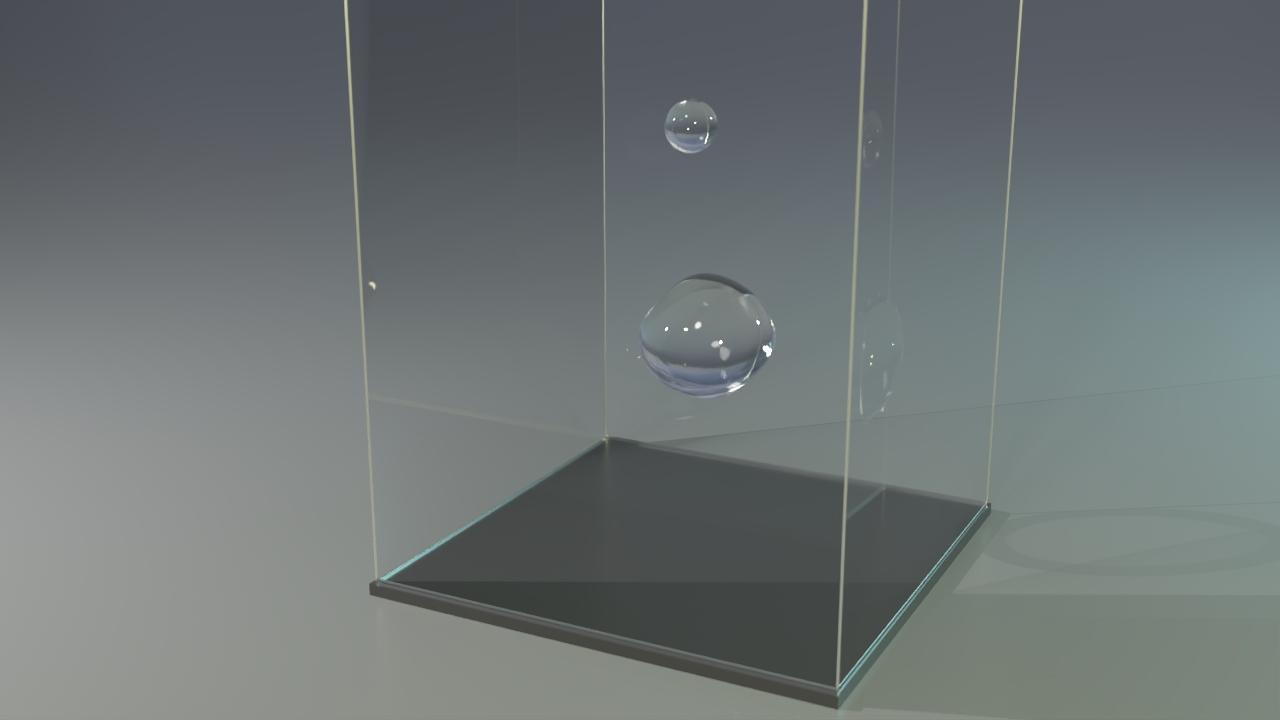}
	\includegraphics[draft=\mydraft,width=0.31\columnwidth,trim={300px 0 300px  0},clip]{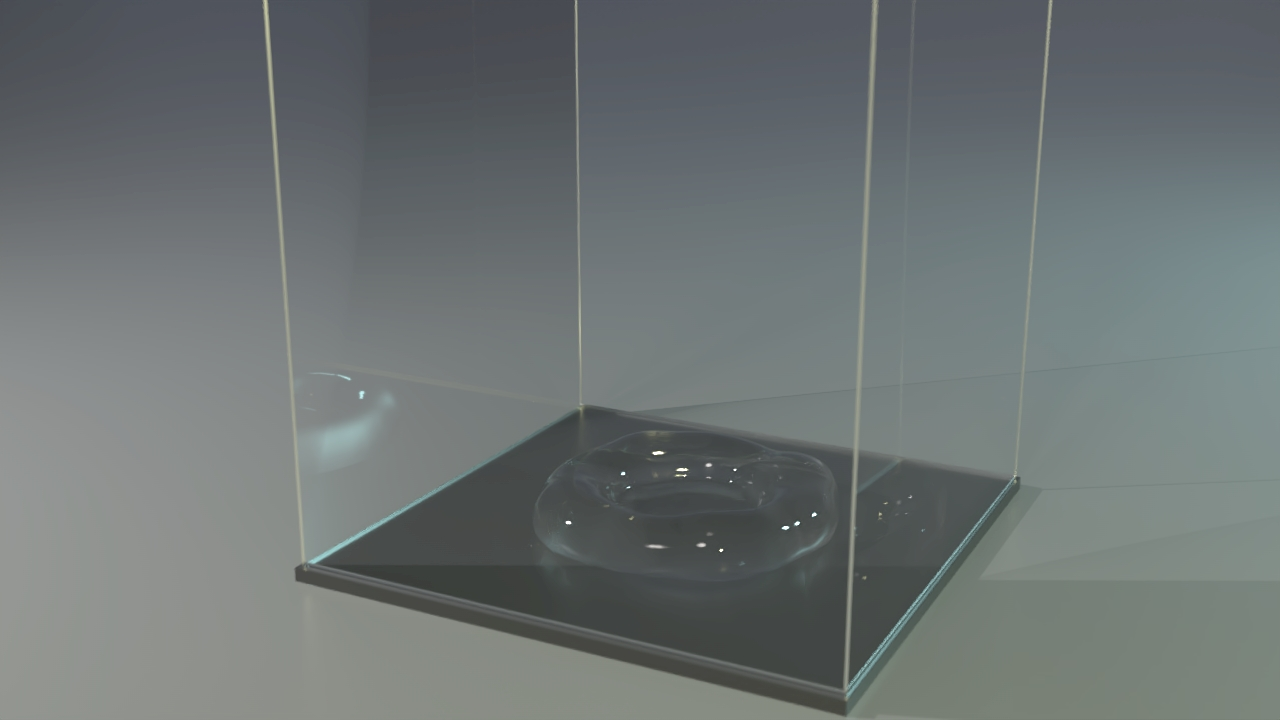}
	\includegraphics[draft=\mydraft,width=0.31\columnwidth,trim={300px 0 300px  0},clip]{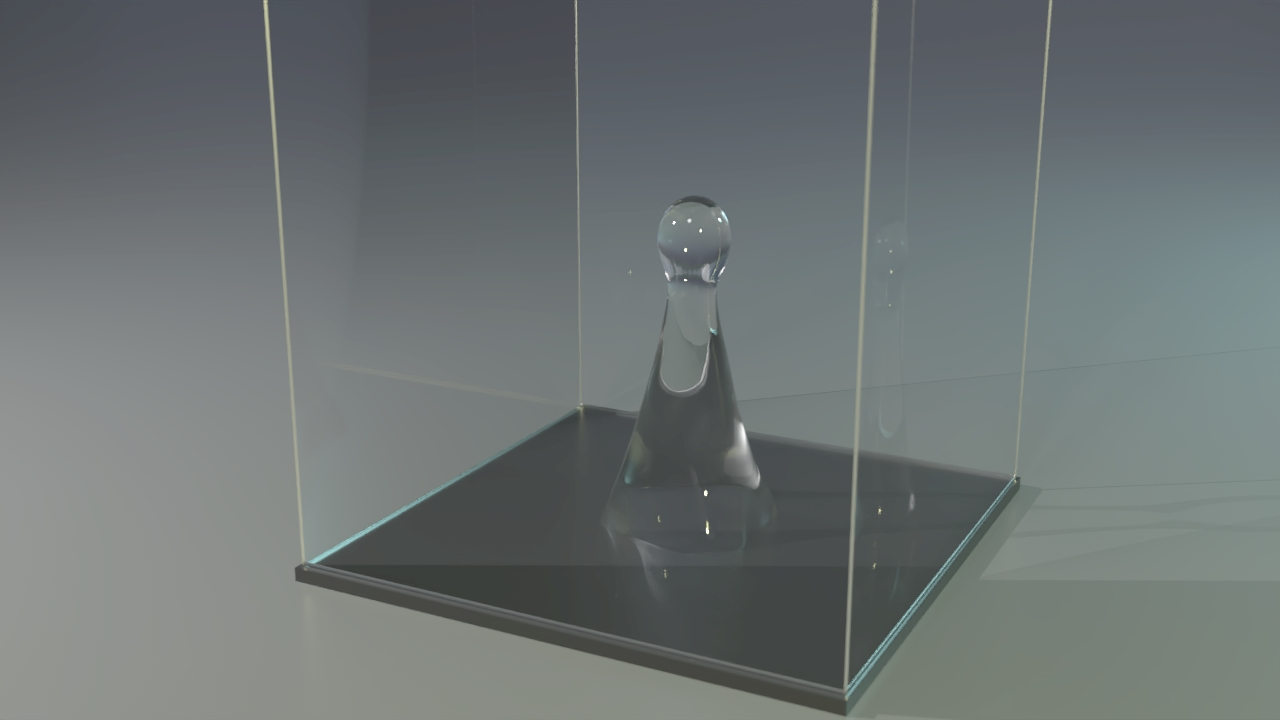}
	\includegraphics[draft=\mydraft,width=0.31\columnwidth,trim={300px 0 300px  0},clip]{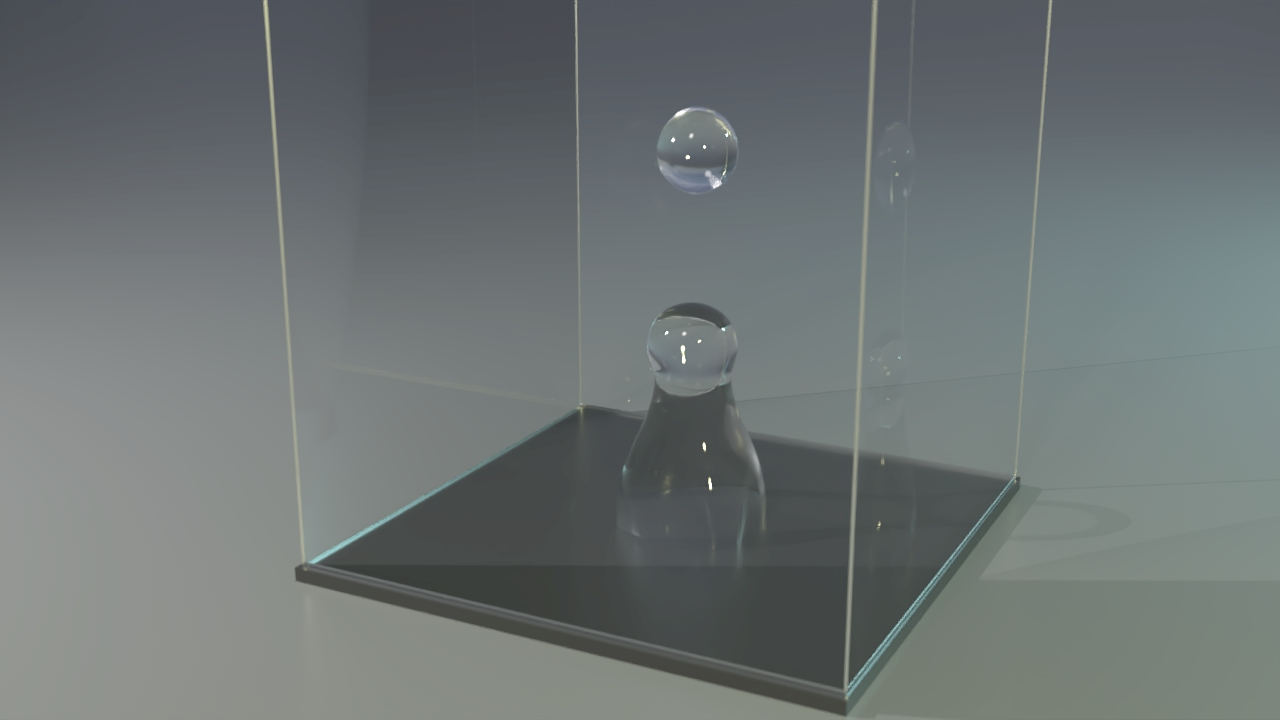}
    \caption{\textit{(Top)} Spherical droplets with different surface tension coefficient free fall from the same height. In the top figure, from left to right, the surface tension coefficients are $k^{\sigma} = 20, 5, 1, 0.1, 0.05$. \textit{(Middle Row)} full rebound of the droplet (initial height: $3.5$ and $k^{\sigma} = 15$). \textit{(Bottom Row)} partial rebound of the droplet (initial height: $2.5$ and $k^{\sigma} = 5$).}
    \label{fig:impact_drop_comparison}
\end{figure}

We demonstrate our method's ability to handle highly dynamic simulations with a wide range of surface tension strengths. 
We simulated several spherical droplets with the same material parameters, each of which free falls from a fixed height and impacts a dry, frictionless, hydrophobic surface. 
The bulk modulus is 83333.33 and the gravitational acceleration is 9.8. 
The same grid resolution of $\Delta x = 1/127$ was used for all the simulations, and the time step was restricted between $10^{-2}$ to $5\times 10^{-5}$ seconds by the CFL condition. 
With different surface tension coefficient $k^{\sigma}$, the droplet showed distinct behaviors upon impact, as shown in Figure \ref{fig:impact_drop_comparison}.
\\
\\
We also captured the partial rebound and the full rebound behaviors of the droplet after the impact. 
The middle and bottom rows of Figure \ref{fig:impact_drop_comparison} show the footage of a droplet with $k^{\sigma} = 15$ dropped from a height of $3.5$ and a droplet with $k^{\sigma} = 5$ dropped from a height of $2.5$, respectively.
With a higher surface tension coefficient and a higher impact speed, the droplet is able to completely leave the surface after the impact. 
Our results qualitatively match the experiment outcomes from Rioboo et al. \shortcite{rioboo:2001:outcomes}.

\subsection{Droplets on Ramps}

\begin{figure}[ht]
    \includegraphics[draft=\mydraft,width=1.0\columnwidth]{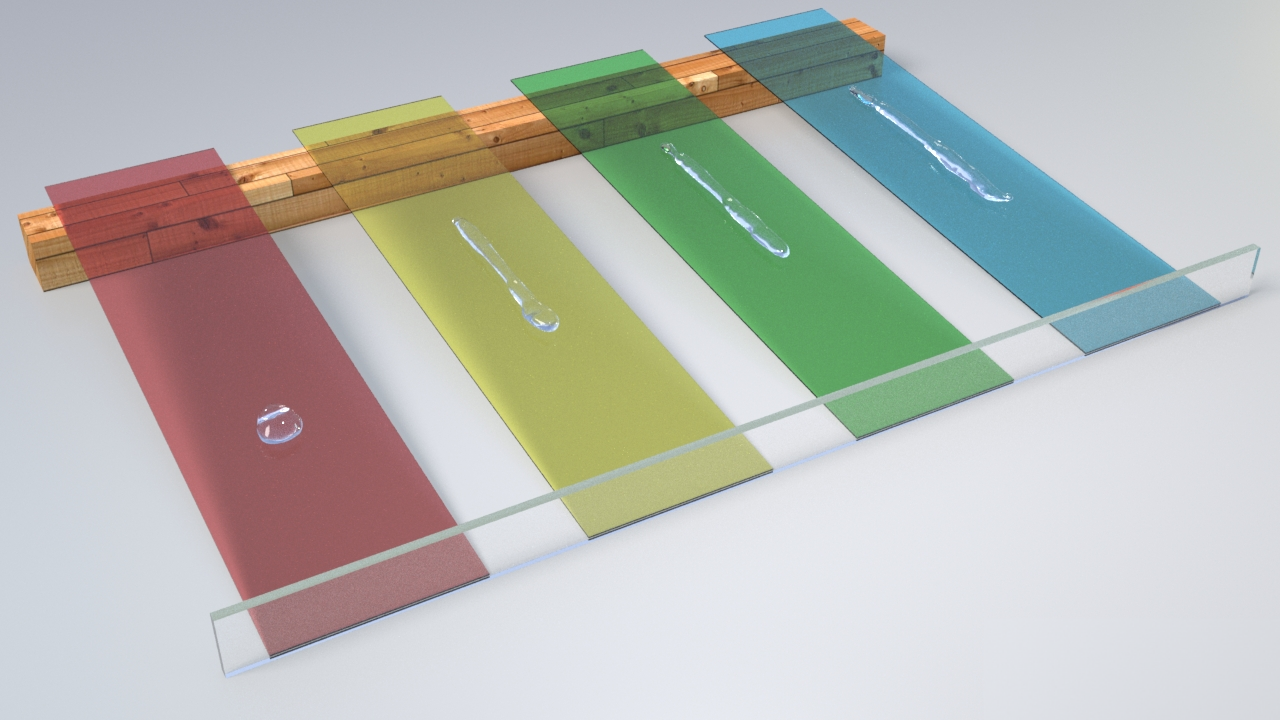}
    \includegraphics[draft=\mydraft,width=1.0\columnwidth]{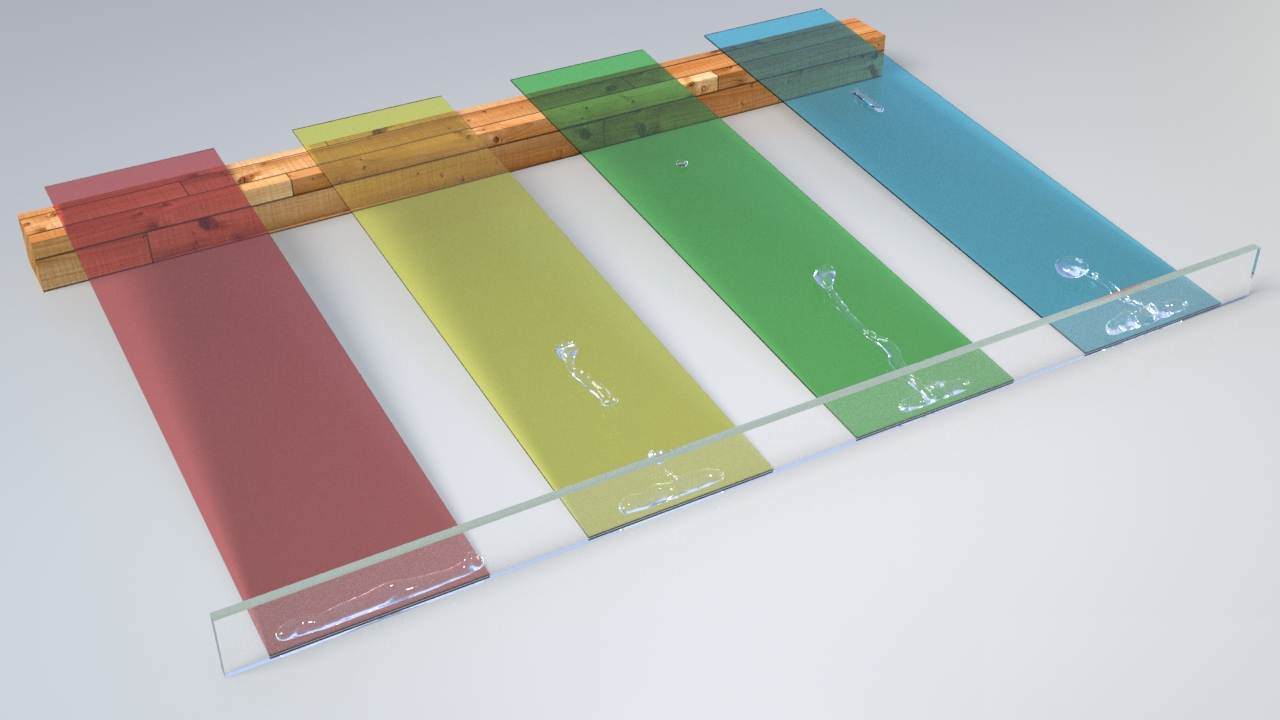}
    \caption{Liquid drops fall on a ramp with varying ratios between the solid-liquid and liquid-air surface tension coefficients.  From left to right: ratios of $0.05$, $0.3$, $0.6$, $1.0$.  \textit{(Top)} Frame 60.  \textit{(Bottom)} Frame 100.}
    \label{fig:ramps}
\end{figure}

As discussed in Section \ref{sec:contact_angle}, our method allows for distinct $k^\sigma$ values at solid-liquid and liquid-air interfaces.
Tuning the ratio between $k^\sigma$ at these interfaces allows simulating different levels of hydrophilicity/hydrophobicity.
Figure \ref{fig:ramps} shows an example of several liquid drops with different $k^\sigma$ ratios falling on ramps of $5.5^{\circ}$ angle.
The grid resolution was $\Delta x = 1/63$, and the time step was restricted between $10^{-2}$ and $10^{-4}$ seconds by the CFL condition.
Coulomb friction with a friction coefficient of $0.2$ was used for the ramp surface.
When there is a larger difference between solid-liquid and liquid-air surface tension coefficients (i.e., a smaller $k^\sigma$ ratio), the liquid tends to drag more on the surface and undergo more separation and sticking.
The leftmost example, with a $k^\sigma$ ratio of $0.05$, exhibits hydrophobic behavior.

\subsection{Lid-Driven Cavity}

\begin{figure}[ht]
    \centering
    \includegraphics[draft=\mydraft,width=0.8\columnwidth]{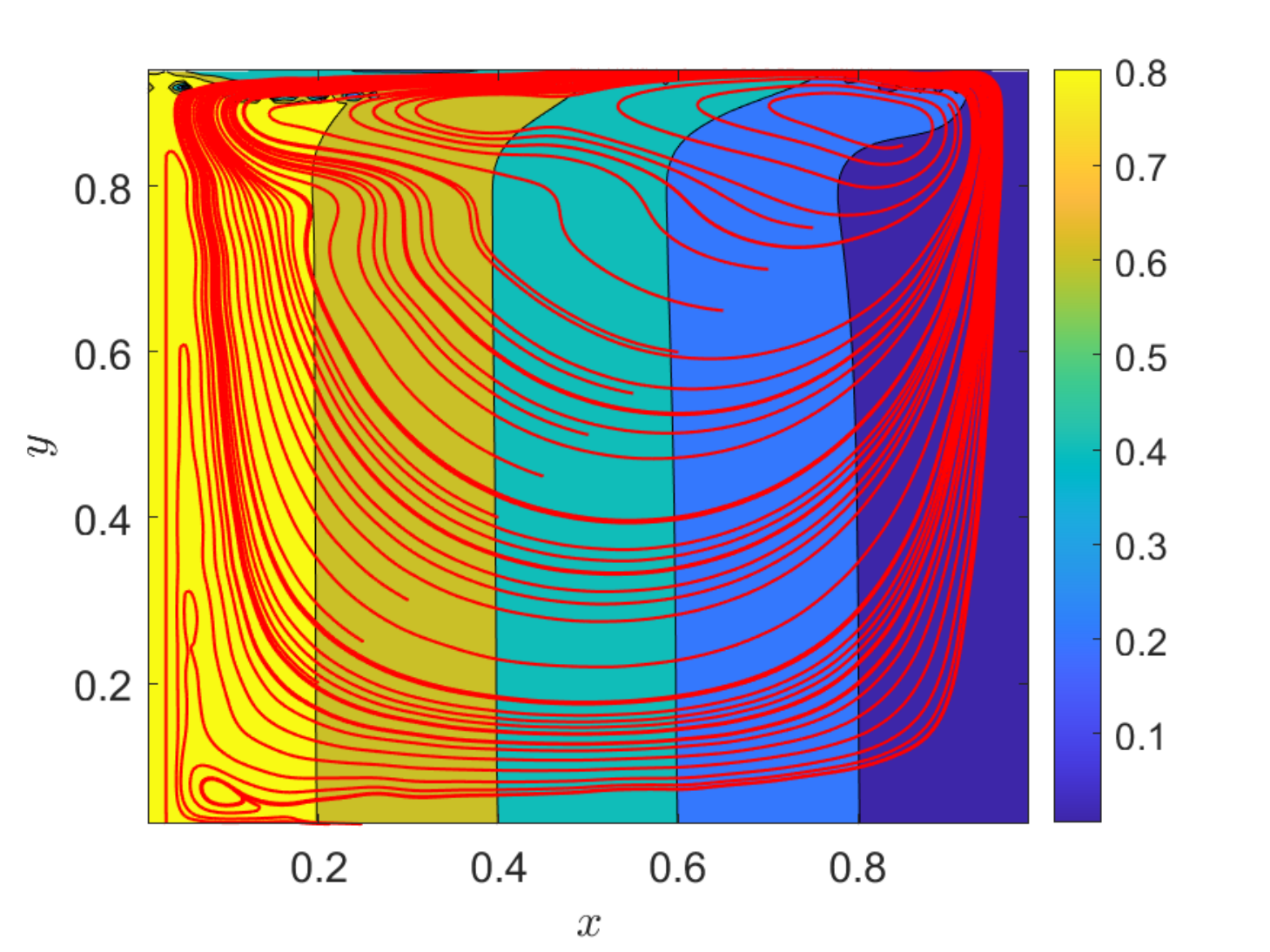}
    \caption{Frame 500 of a two-dimensional lid-driven cavity simulation.  The simulation is initially stationary, but velocity streamlines (red) show the flow pattern characteristic of Marangoni convection that develops due to a temperature-dependent surface tension coefficient.  The contour plot shows the evolving temperature field (initially a linear horizontal distribution).}
    \label{fig:lid-driven-cavity}
\end{figure}

The two-dimensional lid-driven cavity is a classic example in the engineering literature of the Marangoni effect \cite{francois:2006:modeling,hopphirschler:2018:thermocapillary}.
Inspired by works like these, we simulate a square unit domain and fill the domain with particles up to height $1-4\Delta x$ ($\Delta x = 1/63$), which results in a free surface near the top of the domain.
A linear temperature gradient from 1 on the left to 0 on the right is initialized on the particles.
To achieve the Marangoni effect, the surface tension coefficient $k^\sigma$ is set to depend linearly on temperature: $k^\sigma = 1-T_p$.
$k^\sigma$ is clamped to be in $[0,1]$ to avoid artifacts due to numerical precision.
Gravity is set to zero, dynamic viscosity is set to \num{1e-6} and implicit MPM is used with a maximum $\Delta t$ of $0.001$.
Results are shown in Figure \ref{fig:lid-driven-cavity}.
We note that the center of the circulation drifts to the right over the course of the simulation due to uneven particle distribution resulting from the circulation of the particles; investigating particle reseeding strategies to stabilize the flow is interesting future work.

\subsection{Contact Angles}

\begin{figure}[ht]
	\includegraphics[draft=\mydraft,width=0.49\columnwidth,trim={300 175px 300  210px},clip]{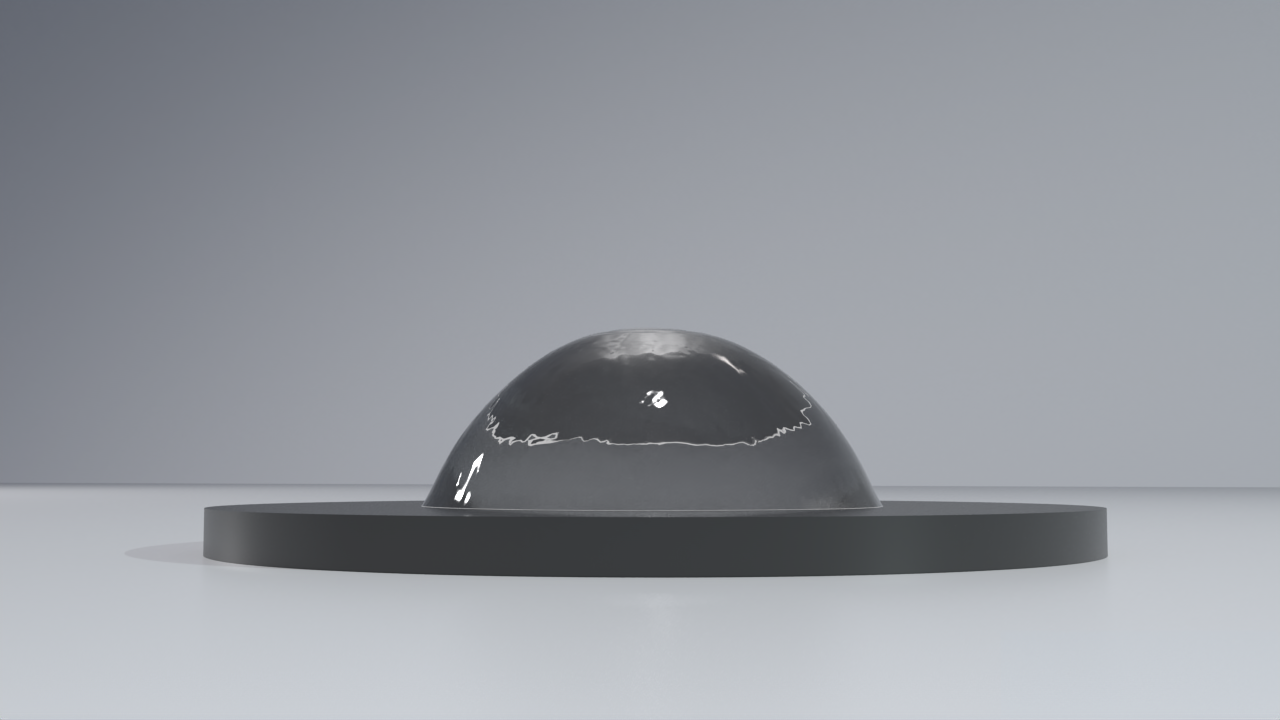}
	\includegraphics[draft=\mydraft,width=0.49\columnwidth,trim={300 175px 300 210px},clip]{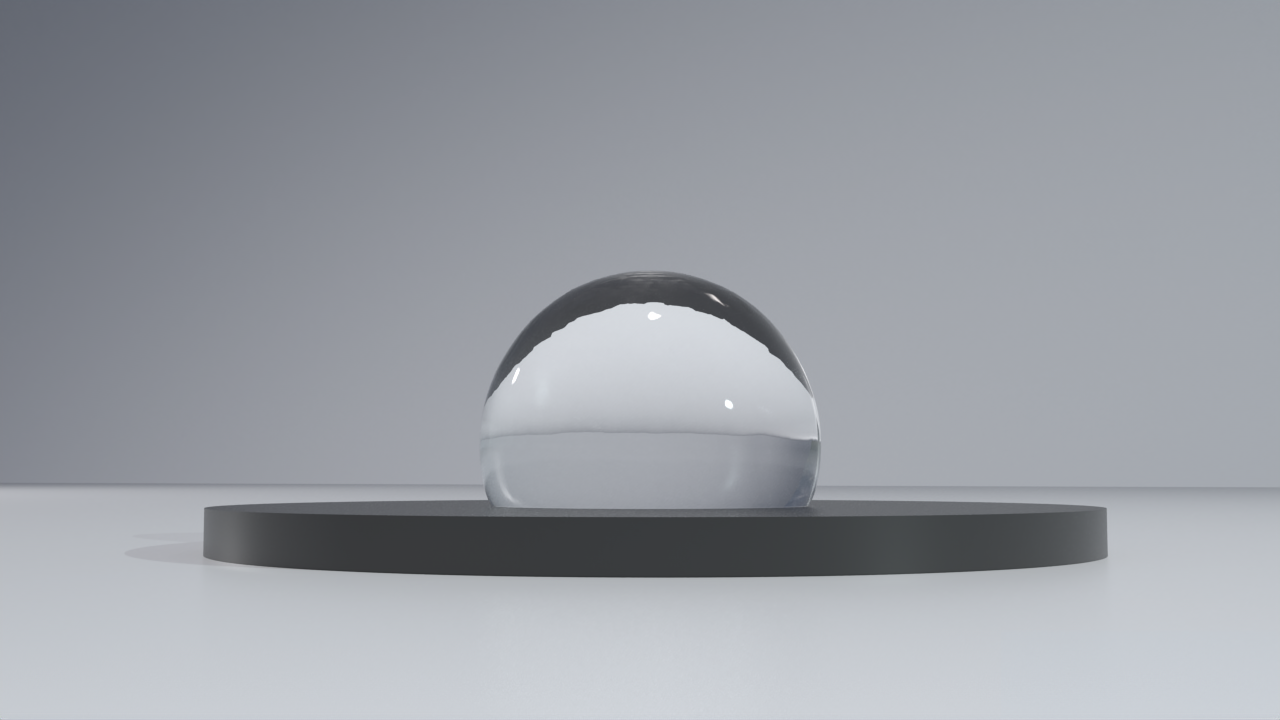}
	\includegraphics[draft=\mydraft,width=0.49\columnwidth,trim={300 175px 300 210px},clip]{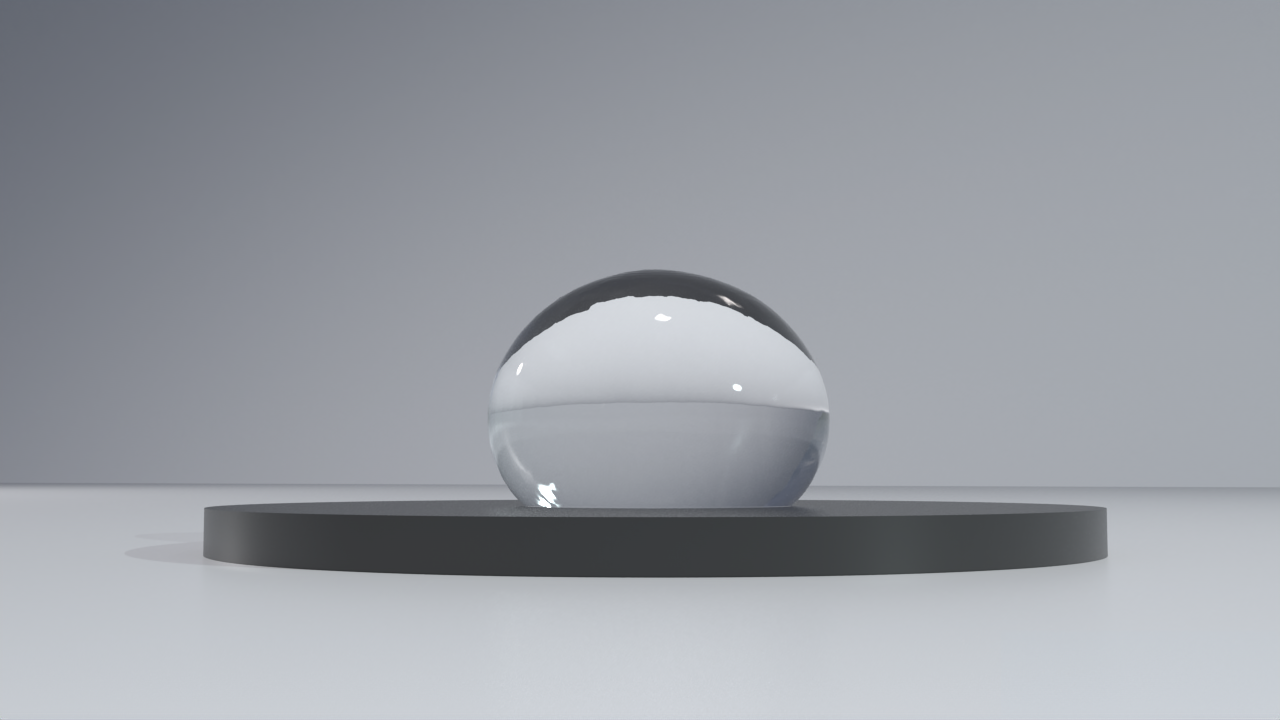}
	\includegraphics[draft=\mydraft,width=0.49\columnwidth,trim={330 175px 270 210px},clip]{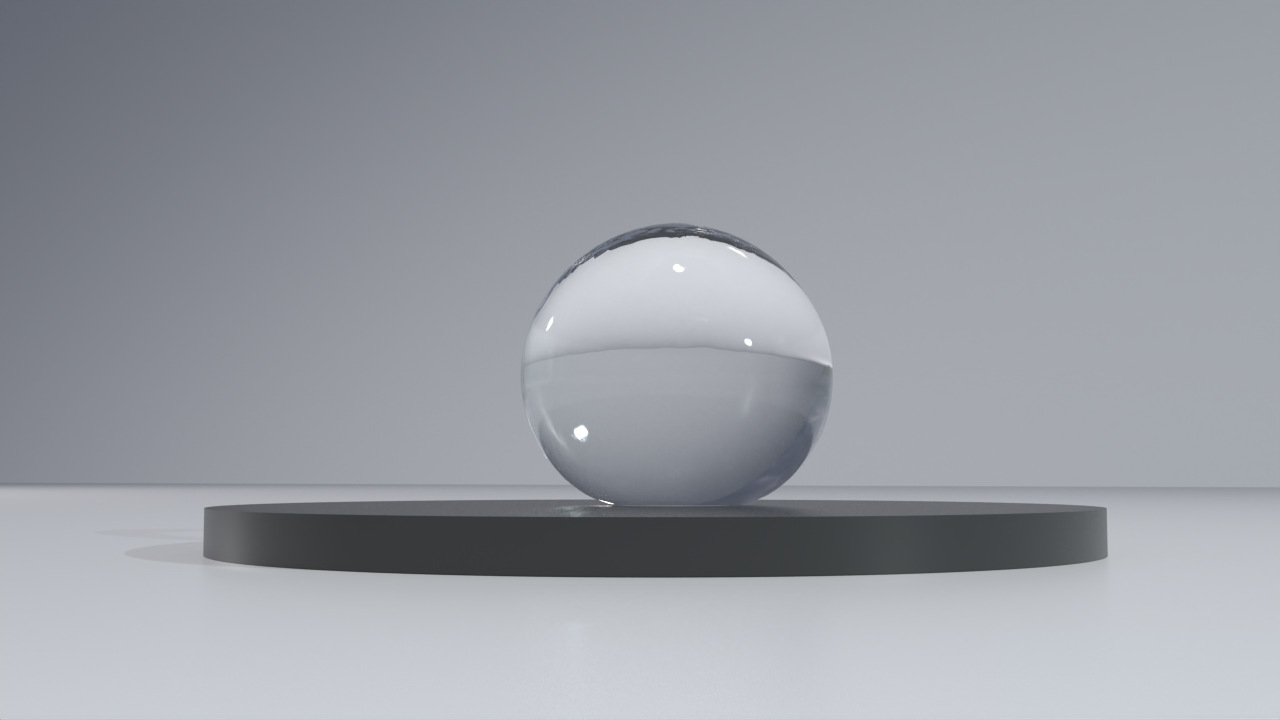}
	\caption{As our droplets settle, we are able to obtain contact angles of approximately 45, 90, 135 and 180 degrees, using a $k^\sigma_{SL}/k^\sigma_{LG}$ ratio of $-\sqrt{2}/2$, $0$, $\sqrt{2}/2$ and $1$, respectively.}
	\label{fig:contact-angles}
\end{figure}

Figure \ref{fig:contact-angles} shows that our method enables simulation of various contact angles, emulating various degrees of hydrophobic or hydrophilic behavior as a droplet settles on a surface. We adjust the contact angles by assigning one surface tension coefficient, $k^\sigma_{LG}$, to the surface particles on the liquid-gas interface, and another one, $k^\sigma_{SL}$, to those on the solid-liquid interface. Following the Young equation (Equation~\eqref{eqn:young}) and our assumption that $k^\sigma_{SG}$ is negligible, the contact angle is given by $\theta = \arccos\big(-k^\sigma_{SL}/k^\sigma_{LG}\big)$. Note that the effect of gravity will result in contact angles slightly smaller than targeted. 
The grid resolution was set to $\Delta x = 1/127$, and the time step restricted between $0.0333$ and $1\times10^{-4}$. Each droplet is discretized using $230$k interior particles and $250$k surface particles. The bulk modulus of the fluid is $83333.33$, $k^\sigma_{LG}$ is set to $2$, and the dynamic viscosity is $0.075$.

\subsection{Soap Droplet in Water}

\begin{figure}[ht]
    \includegraphics[draft=\mydraft,width=.49\columnwidth,trim={200px 0 200px 0},clip]{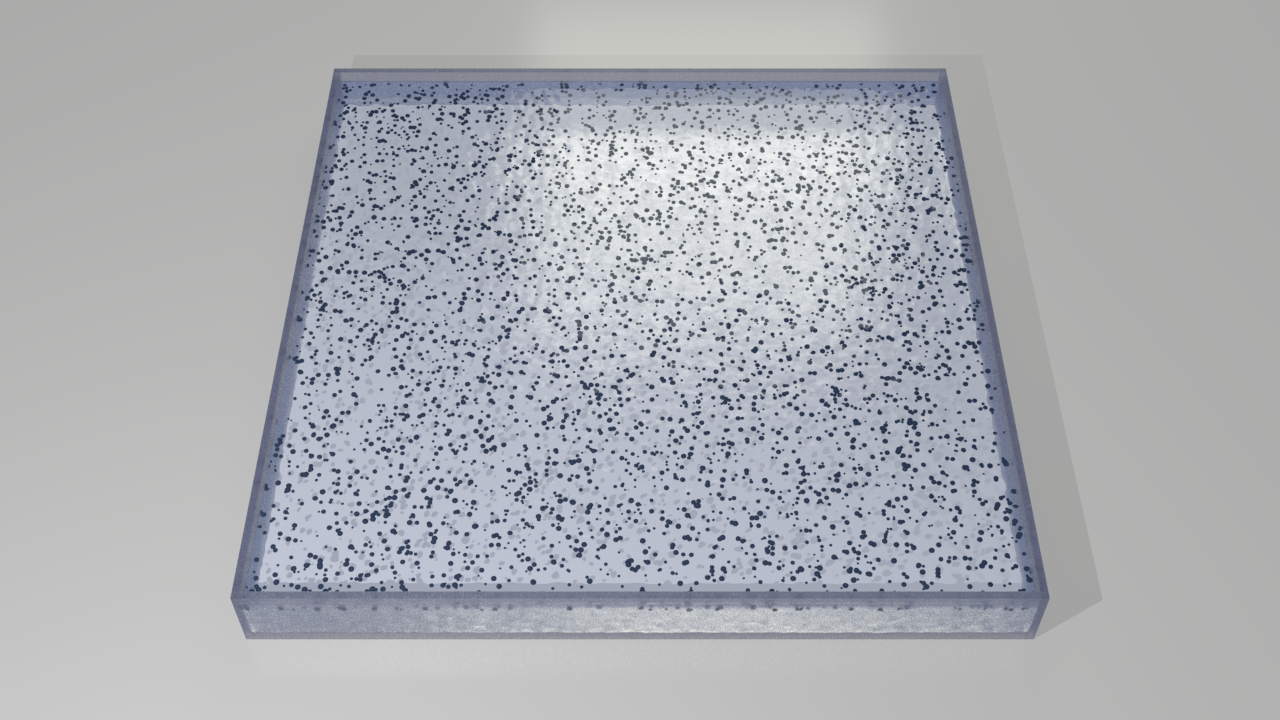}
    \includegraphics[draft=\mydraft,width=.49\columnwidth,trim={200px 0 200px 0},clip]{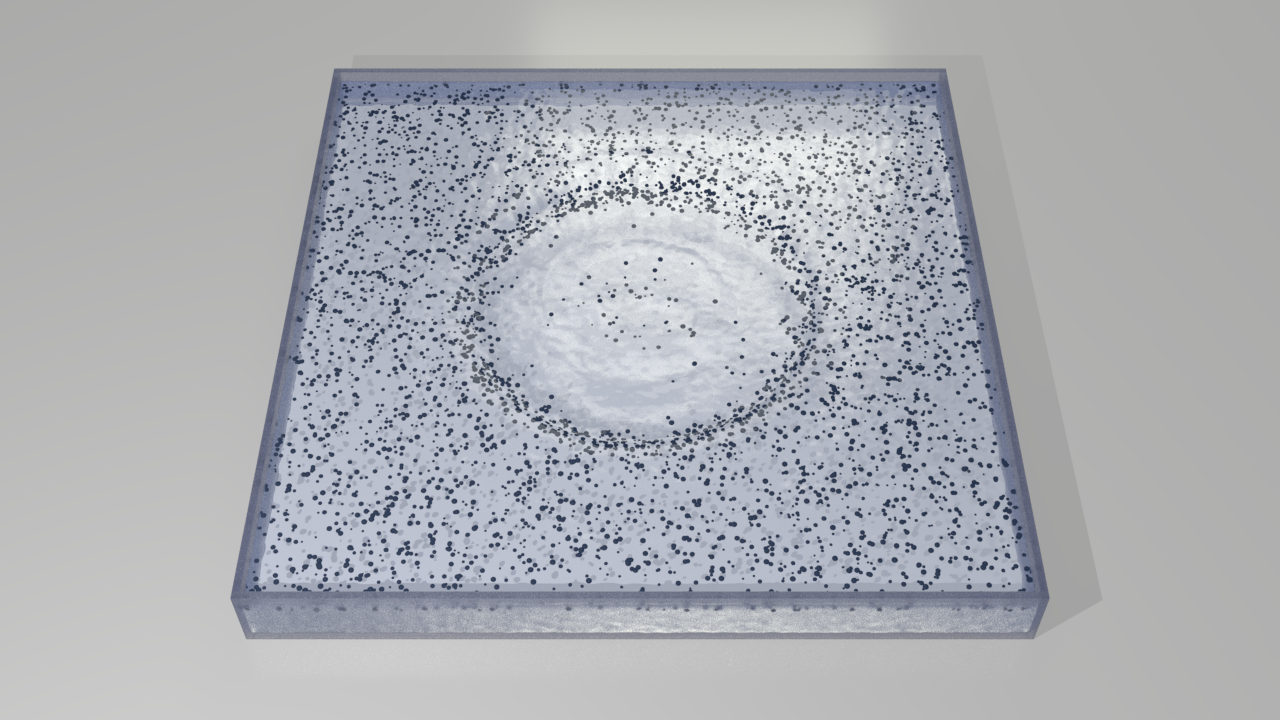}
    \includegraphics[draft=\mydraft,width=.49\columnwidth,trim={200px 0 200px 0},clip]{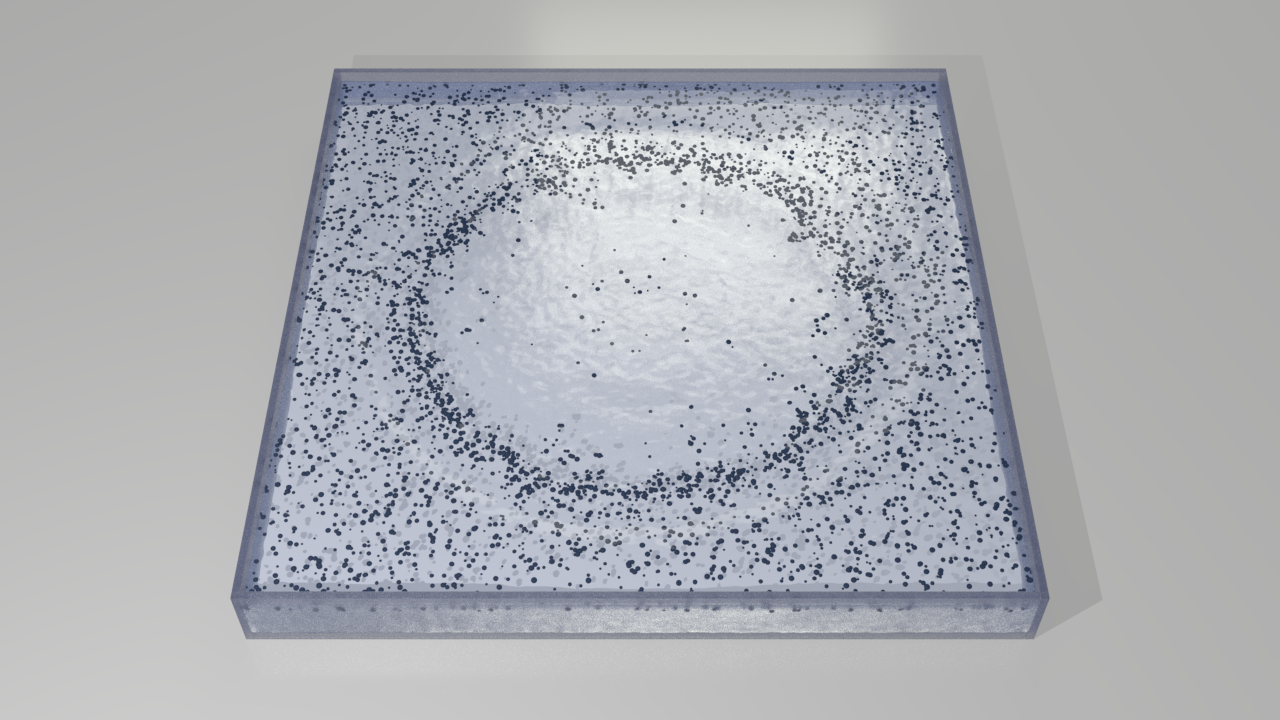}
    \includegraphics[draft=\mydraft,width=.49\columnwidth,trim={200px 0 200px 0},clip]{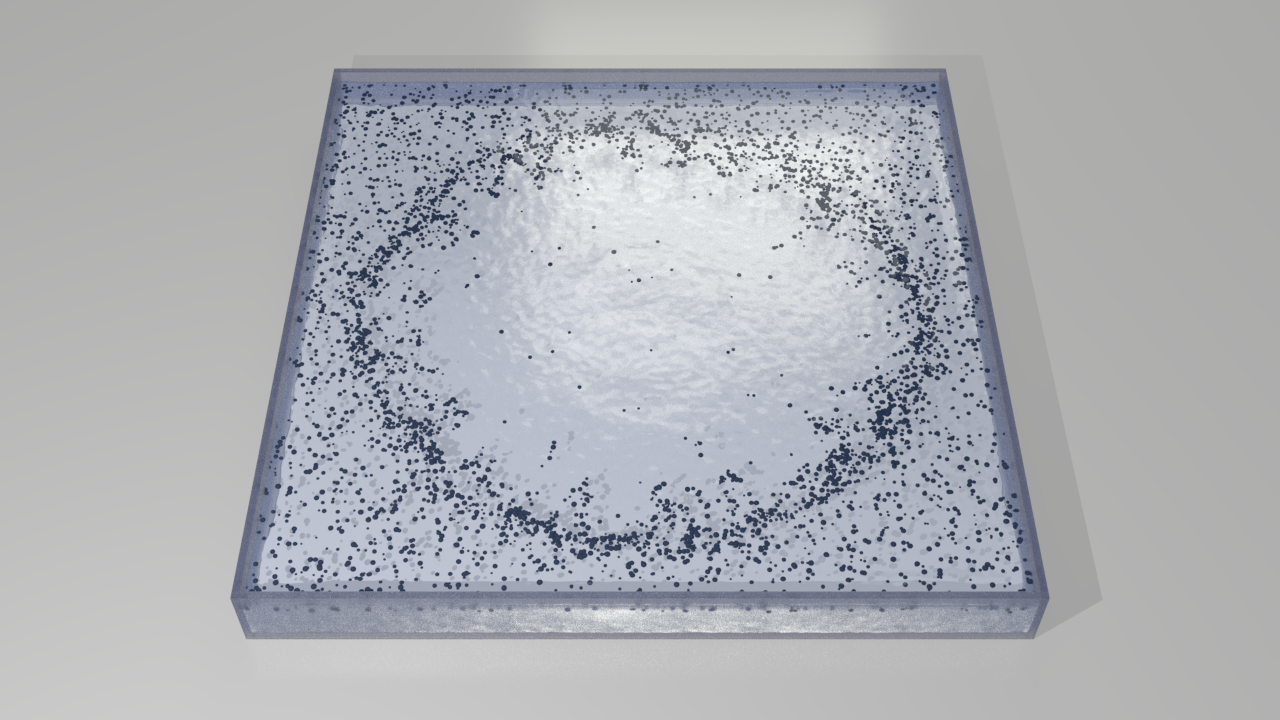}
    \caption{The soap in the center of the pool surface reduces the surface tension. The surface tension gradient drives the markers towards the walls of the container. Frames 0, 10, 20, 40 are shown in this footage.}
    \label{fig:soap-water}
\end{figure}

We demonstrate a surface tension driven flow by simulating the soap reducing the surface tension of the water. 
We initialize a $1\times 0.05 \times 1$ rectangular water pool and identify the particles in the middle of the top surface as liquid soap. 
Boundary particles associated with the water particles have higher surface tension than those associated with the soap particles.
In order to visualize the effect of the surface tension driven flow, we randomly selected marker particles on the top surface of the pool.
Due to the presence of the soap, the center of the pool has lower surface tension than the area near the edge of the container.
Figure \ref{fig:soap-water} shows footage of this process.
A grid resolution of $\Delta x = 1/127$ was used, and the time step was limited between $10^{-2}$ and $5\times 10^{-5}$ by the CFL number.
The bulk modulus of water is set to be $16666.67$. The surface tension $k^{\sigma}$ for water is set to be $0.5$ and $k^{\sigma}$ for soap is $0.01$.

\subsection{Wine Glass}

\begin{figure}[ht]
    \includegraphics[draft=\mydraft,width=.49\columnwidth,trim={270px 0 250px 0},clip]{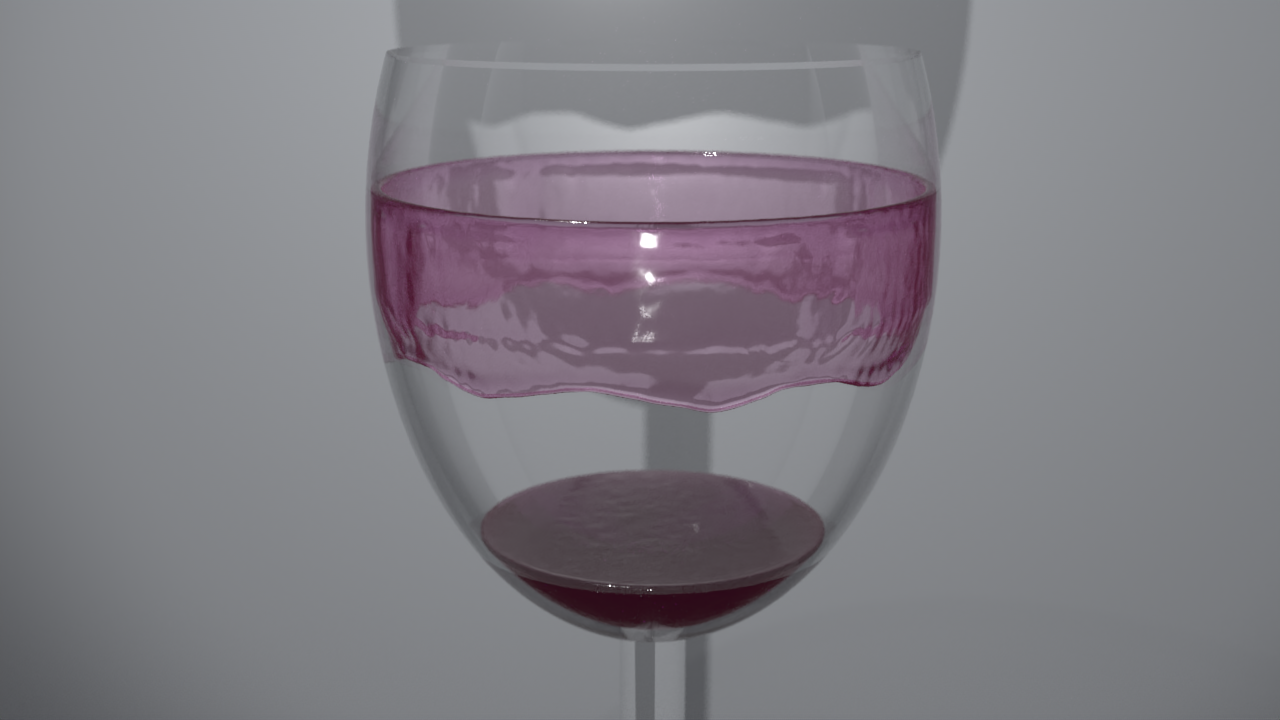}
    \includegraphics[draft=\mydraft,width=.49\columnwidth,trim={270px 0 250px 0},clip]{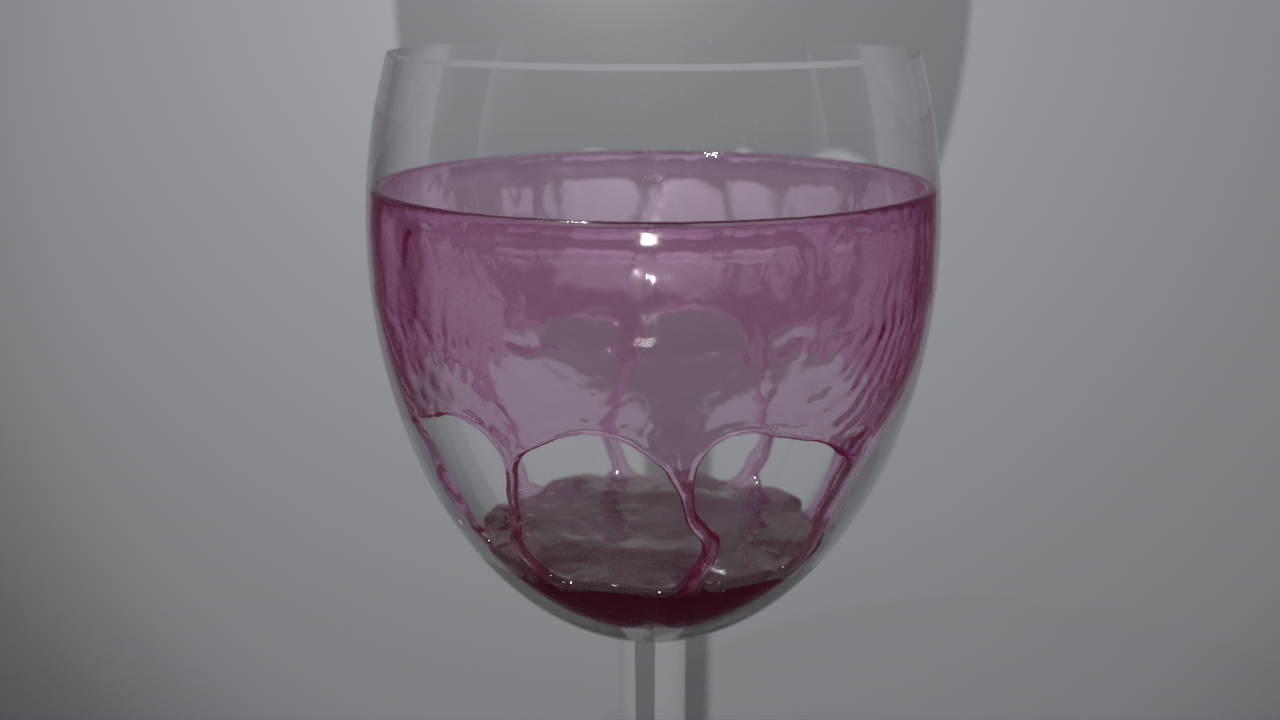}
    \caption{Wine is initialized in a glass with part of the interior pre-wetted.  The falling wine forms tears and ridges, and the tears eventually connect with the bulk fluid.  Frames 30 and 90 are shown.}
    \label{fig:wine-glass}
\end{figure}

We consider an example of wine flowing on the surface of a pre-wetted glass.
The glass is an ellipsoid centered at $(0.5, 0.7, 0.5)$ with characteristic dimensions $a = 0.4$, $b = 0.6$, and $c = 0.4$.
We initialize a thin band of particles with thickness of $2\Delta x$ on the surface of the wine glass and observe the formation of ridges and fingers as the particles settle toward the bulk fluid in the glass.
The grid resolution is $\Delta x = 1/127$ and a maximum allowable time step is $10^{-2}$ constrained by the CFL number.
We set the surface tension coefficient on the liquid-gas interface is $k^{\sigma}_{LG} = 0.05$ and the one on the solid-liquid interface is $k^{\sigma}_{SL} = 0.015$.
The piecewise constant surface tension gradient leads to a more prominent streaking behaviors of the liquid on the glass wall.
The results are shown in Figure \ref{fig:wine-glass}.

\subsection{Candles}

\begin{figure}[ht]
    \includegraphics[draft=\mydraft,width=.49\columnwidth,trim={180px 0 220px 0},clip]{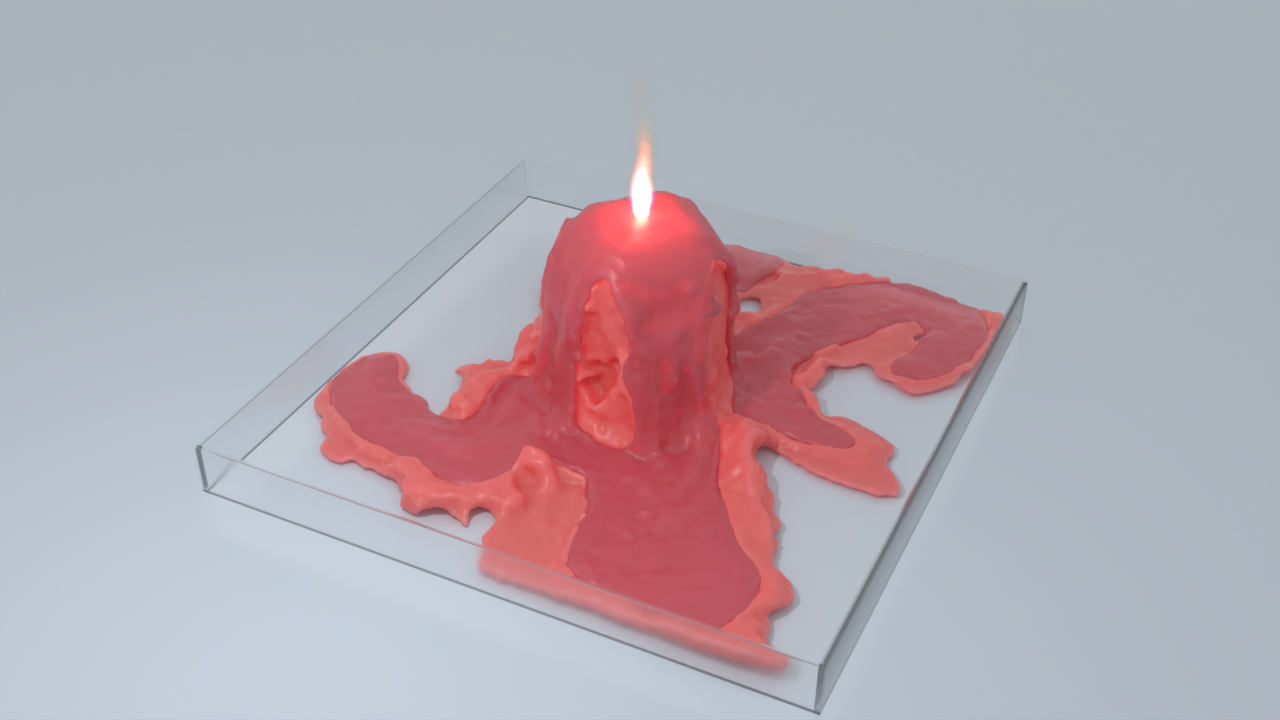}
    \includegraphics[draft=\mydraft,width=.49\columnwidth,trim={270px 0 330px 0},clip]{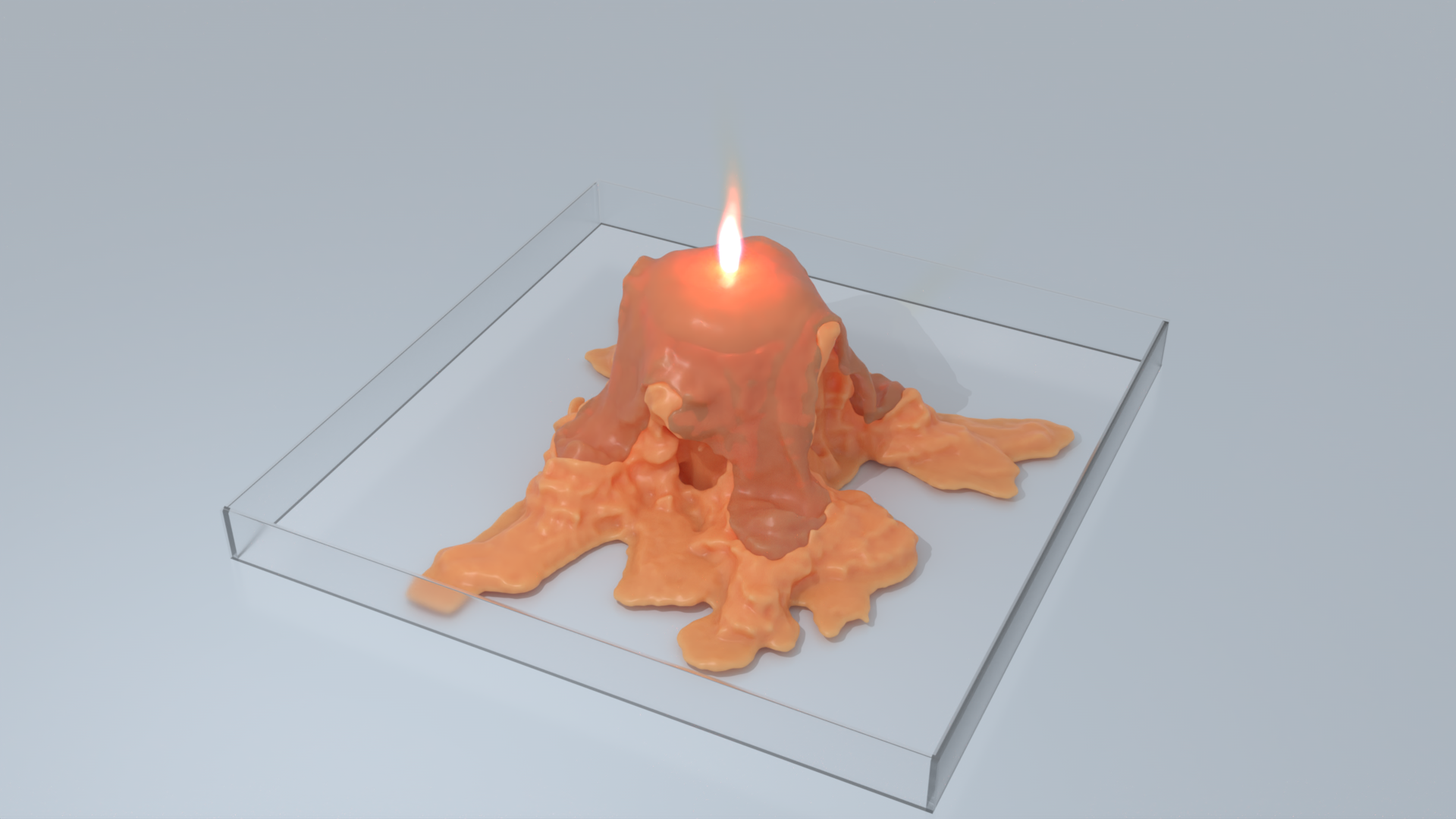}
    \includegraphics[draft=\mydraft,width=.49\columnwidth,trim={180px 0 220px 0},clip]{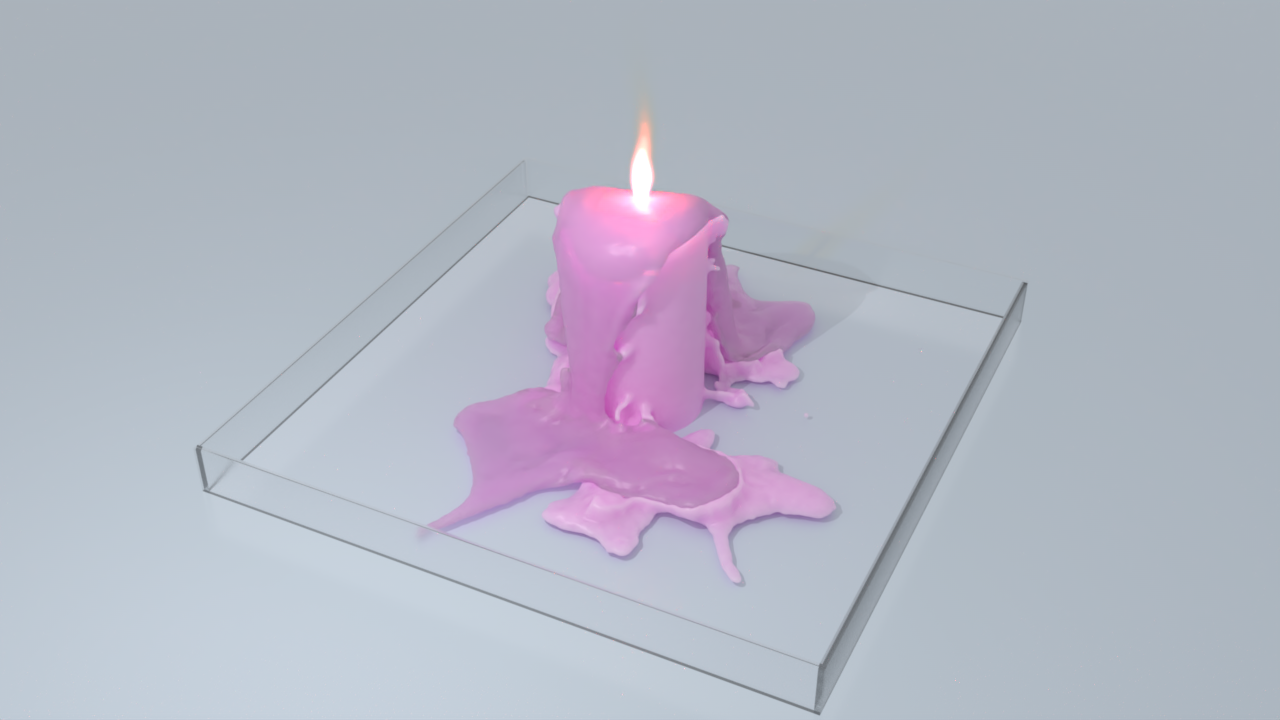}
    \includegraphics[draft=\mydraft,width=.49\columnwidth,trim={270px 0 330px 0},clip]{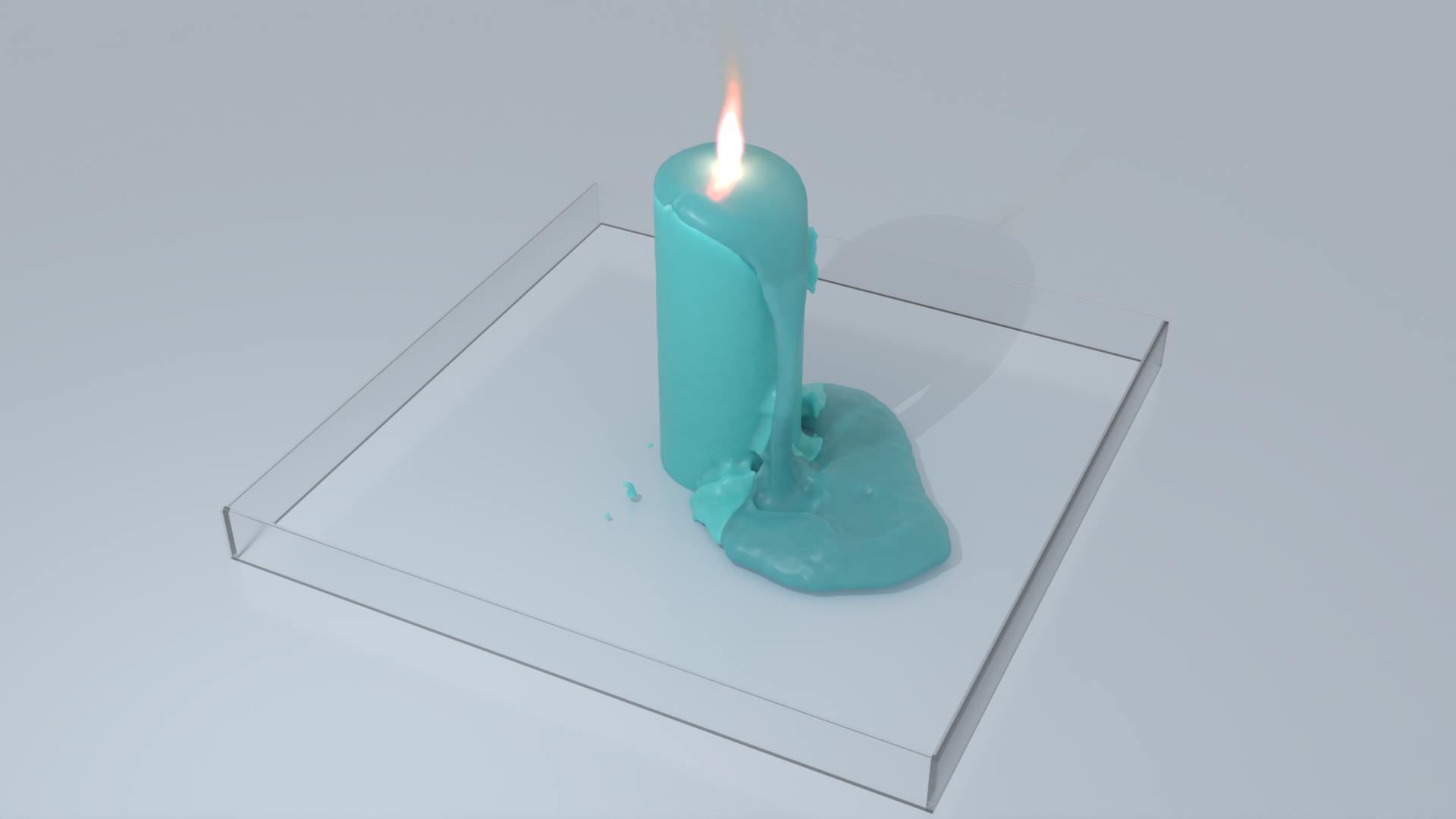}
    \caption{Various $k^\sigma$ values ($0.05$, $0.1$, $0.2$, $0.4$) are simulated in the case of a melting candle.  Frame 1202 is shown.}
    \label{fig:candle-kst-wedge}
\end{figure}

\begin{figure}[ht]
    \includegraphics[draft=\mydraft,width=1.0\columnwidth,trim={0 160px 0 160px},clip]{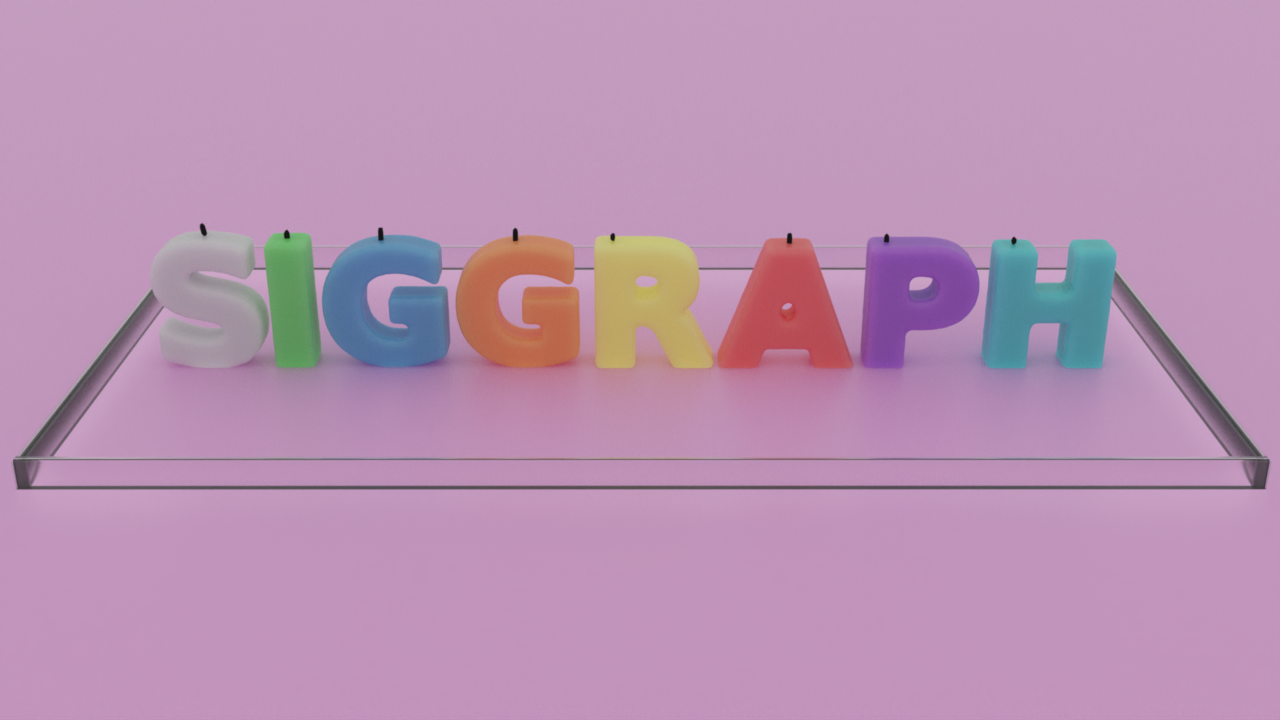}
    \includegraphics[draft=\mydraft,width=1.0\columnwidth,trim={0 160px 0 160px},clip]{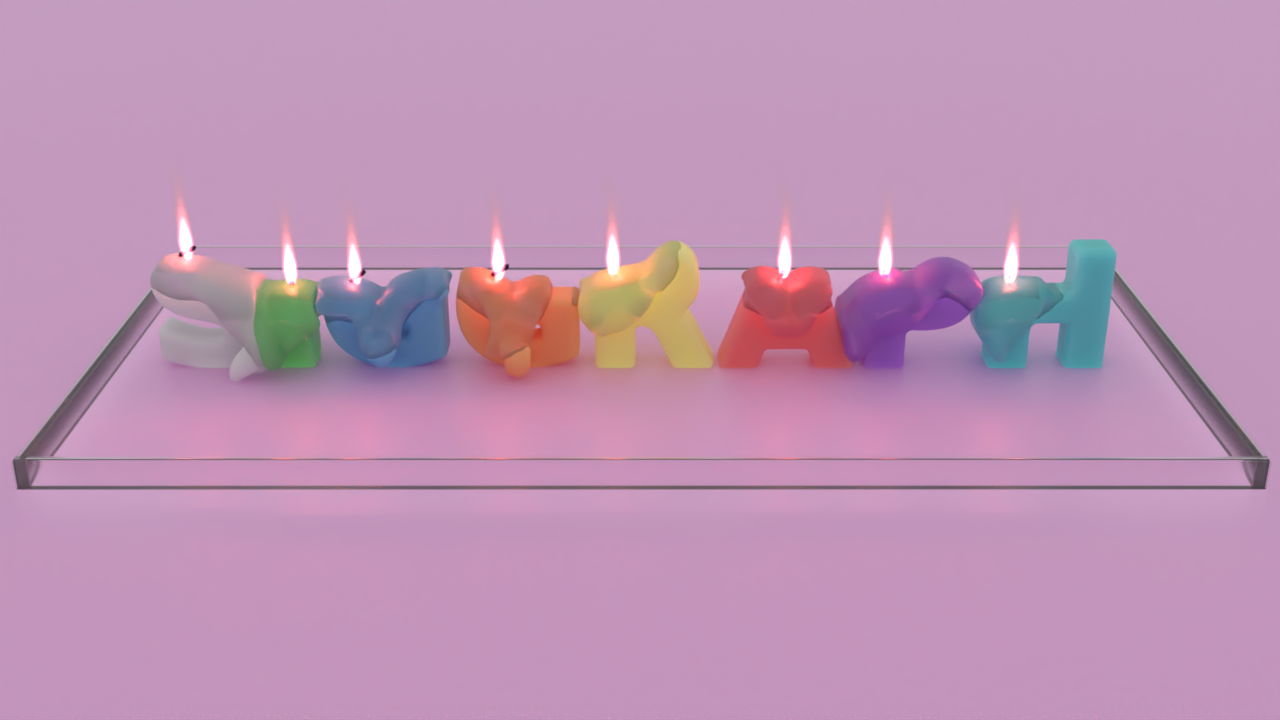}
    \includegraphics[draft=\mydraft,width=1.0\columnwidth,trim={0 160px 0 160px},clip]{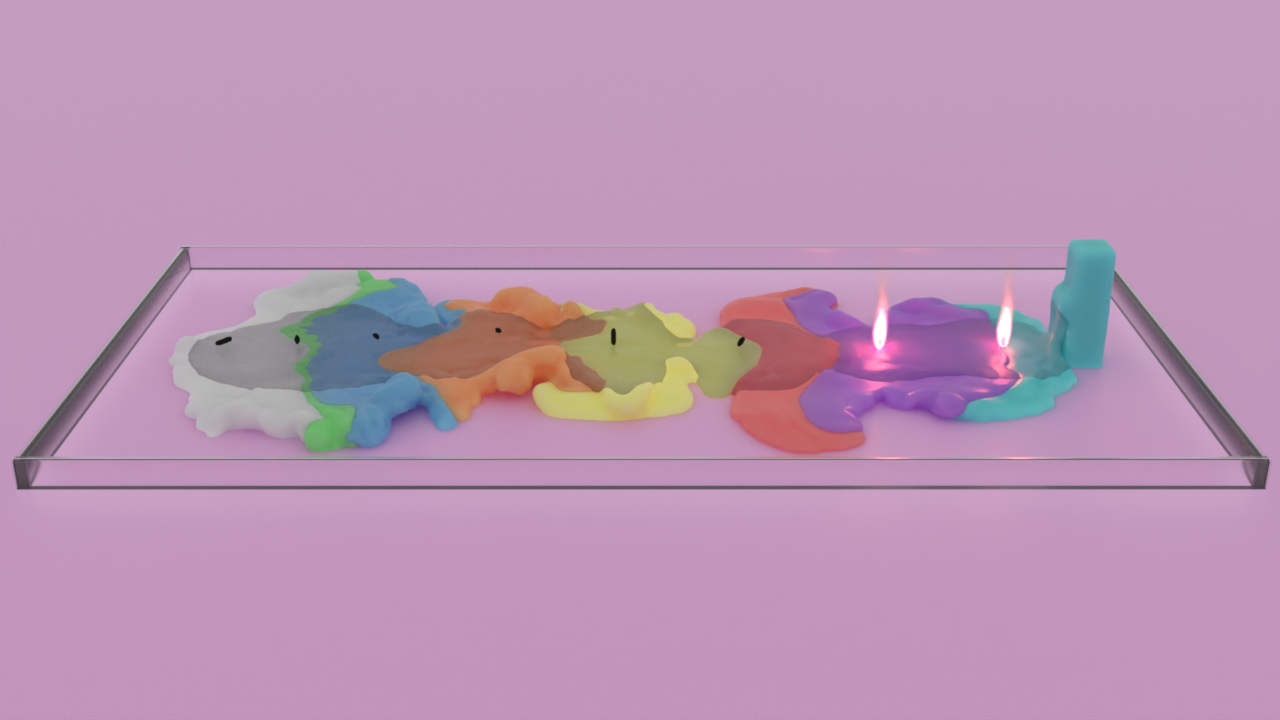}
    \caption{Letter-shaped candles melt inside a container.  \textit{(Top)} Frame 1, before flames are lit.  \textit{(Middle)} Frame 60, in the middle of melting.  \textit{(Bottom)} Frame 200, as flames are extinguished and wax pools resolidify.}
    \label{fig:letters}
\end{figure}

We simulate several scenarios with wax candles.
In these examples, wax melts due to a heat source (candle flame) and resolidifies when it flows away from the flame.
Ambient temperature $\hat T$ is taken to be $298\text{K}$, and the melting point is $303\text{K}$.
Thermal diffusivity $K$ is taken to be $0.1$, and specific heat capacity $c_p$ is set to $1$.
No internal heat source is used ($H = 0$); instead, heating and cooling are applied only via the boundary conditions.\\
\\
To simulate the candle wicks, we manually construct and sample points on cubic splines.
As the simulation progresses, we delete particles from the wick that are too far above the highest ($y$-direction) liquid particle within a neighborhood of the wick.
The flames are created by running a separate FLIP simulation as a postprocess and anchoring the result to the exposed portion of each wick.
We rendered these scenes using Arnold \cite{georgiev:2018:arnold} and postprocessed the renders using the NVIDIA OptiX denoiser (based on Chaitanya et al.\ \shortcite{chaitanya:2017:interactive}).\\
\\
We consider the effect of varying $k^\sigma$ on the overall behavior of the flow.
Figure \ref{fig:candle-kst-wedge} compares $k^\sigma$ values of $0.05$, $0.1$, $0.2$, and $0.4$.
In these examples, a grid resolution of $\Delta x = 1/127$ was used, along with boundary condition parameters $h=0.5$ and $b=50$.
The figure demonstrates that as surface tension increases, the molten wax spreads significantly less.
As the wax cools and resolidifies, visually interesting layering behavior is observed.
\\
Figure \ref{fig:letters} shows an example of several candle letters melting in a container.
Wicks follow generally curved paths inside the letters.
Melt pools from the different letters seamlessly interact.
This simulation used a surface tension coefficient $k^\sigma = 0.05$, $h=2.5$, $b=100$, dynamic viscosity of $0.01$, $\Delta x = 1/127$, and a bulk modulus of $83333.33$ for liquid and solid phases.

\subsection{Droplet with Marangoni Effect}

\begin{figure}
	\includegraphics[draft=\mydraft,width=0.49\columnwidth,trim={200px 150px 600px 300px},clip]{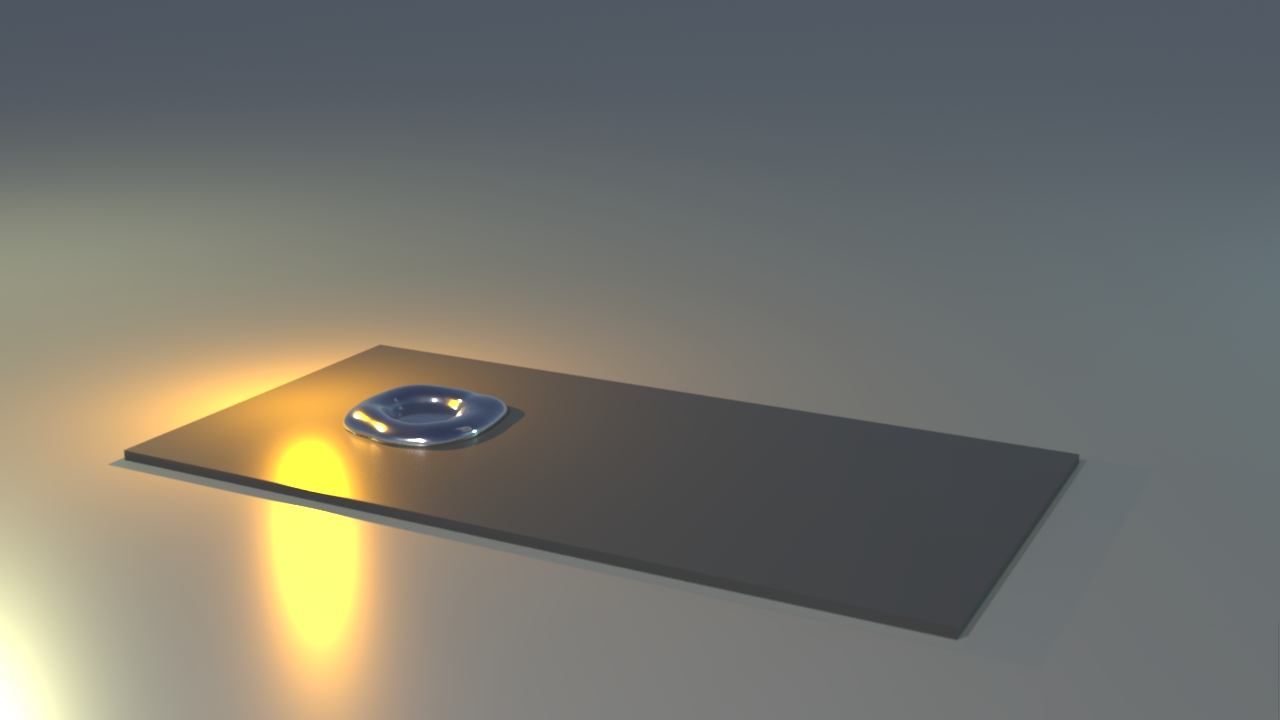}
	\includegraphics[draft=\mydraft,width=0.49\columnwidth,trim={200px 150px 600px 300px},clip]{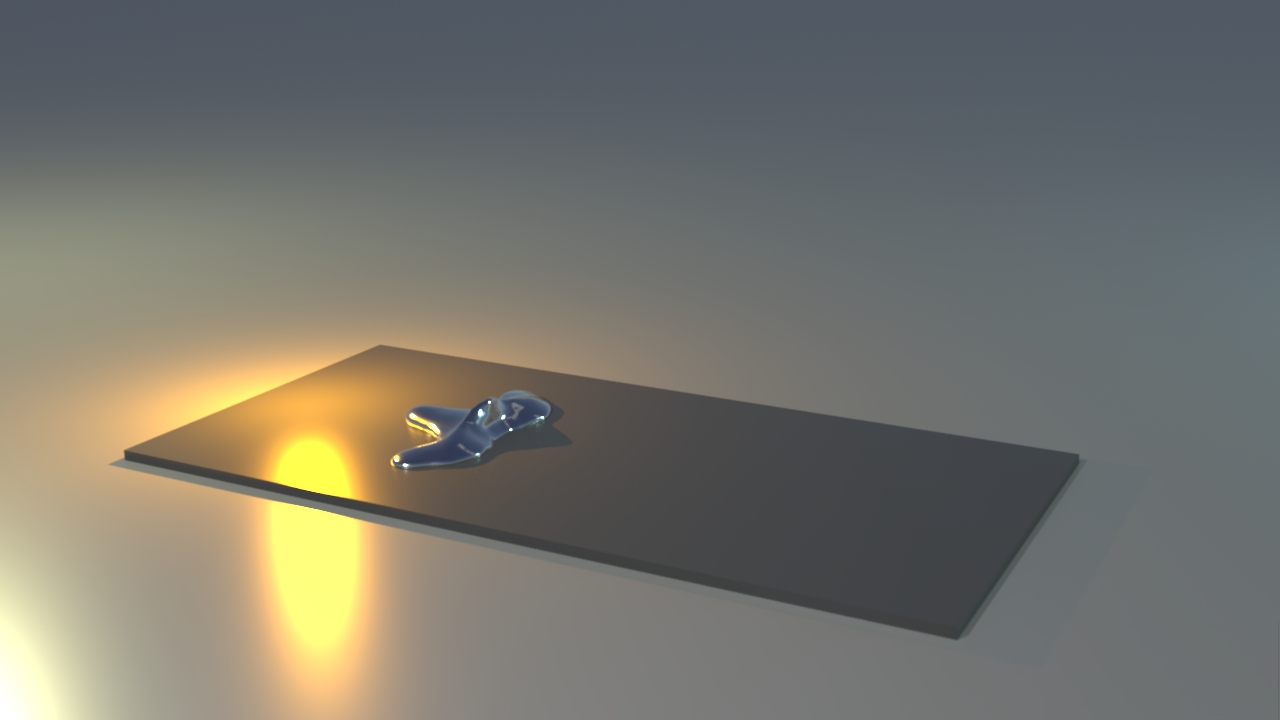}
	\includegraphics[draft=\mydraft,width=0.49\columnwidth,trim={200px 150px 600px 300px},clip]{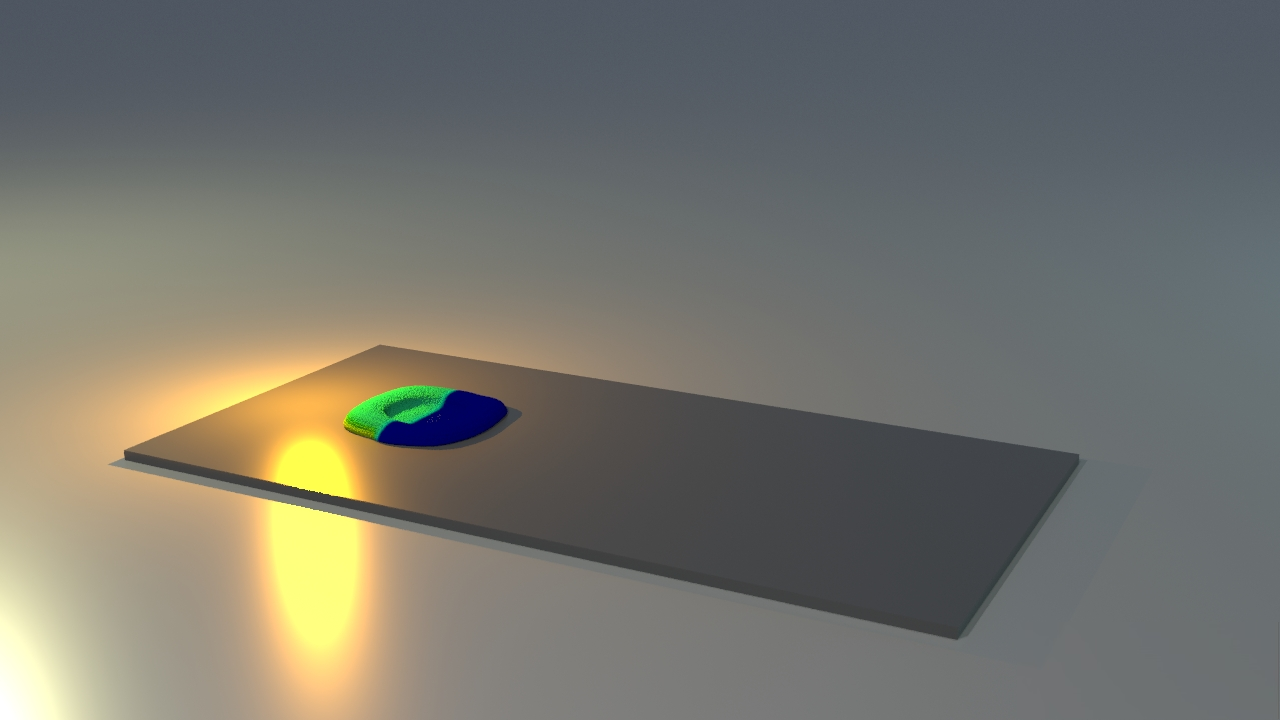}
	\includegraphics[draft=\mydraft,width=0.49\columnwidth,trim={200px 150px 600px 300px},clip]{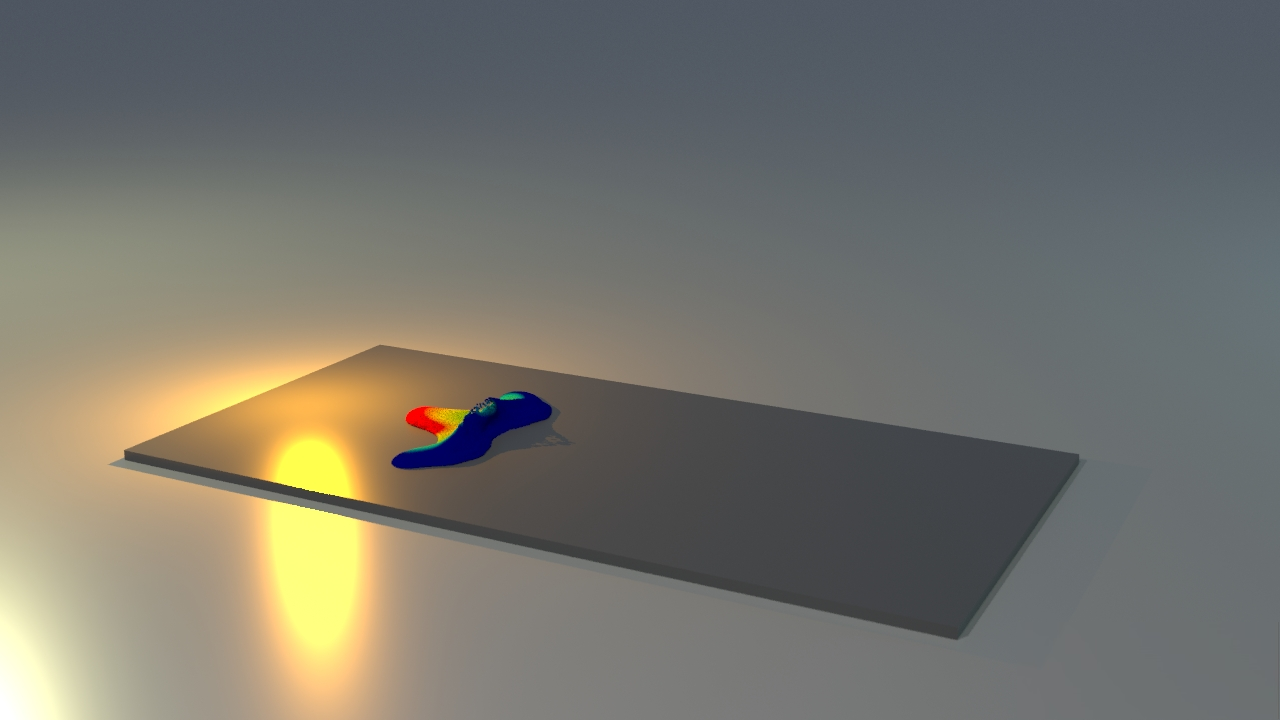}
	\caption{A liquid metal droplet subjected to heating on one side. The surface tension coefficient increases as the temperature increases. \textit{(Top)} the liquid metal at frame 45 and frame 130. \textit{(Bottom)} the particle view of temperature distribution at frame 45 and frame 130. The red color indicates higher temperature.}
	\label{fig:marangoni_droplet}
\end{figure}

We simulated a liquid metal droplet that moves under the Marangoni effect, i.e.\ due to a temperature-induced surface tension gradient. 
The droplet first falls to the ground and spreads on the dry surface. 
We then turn on the heating while the droplet is still spreading; only one side of the droplet is subjected to heating.
At its original temperature, the surface tension coefficient $k^{\sigma}$ of the droplet is $0.5$.
As the temperature increase, $k^{\sigma}$ increases linearly with the temperature.
When the change in temperature is greater than $50K$, the surface tension coefficient reaches its maximum value of $5$.
Since the hotter side of the droplet has higher surface tension, the stronger surface tension drives the particles to flow to the colder side as shown in Figure \ref{fig:marangoni_droplet}.
This surface tension gradient results in an interesting self-propelled behavior of the liquid metal droplet.
A domain of size $2 \times 1 \times 1$ with a grid resolution of $\Delta x = 1/127$ is used for this example.
The time step is restricted between $10^{-2}$ to $5\times 10^{-5}$ seconds by the CFL condition.
The bulk modulus of the liquid is $83333.33$.
$h=0.1$ and $b=50$ are used for the thermal boundary conditions.

\subsection{Performance}

Table \ref{tbl:perf} shows average per-timestep runtime details for several of our examples.  For this table, all experiments were run on a workstation equipped with 128GB RAM and with dual Intel\textsuperscript{\textregistered} Xeon\textsuperscript{\textregistered} E5-2687W v4 CPUs at 3.00Ghz.

\begin{table}[tbp]
\caption{Performance measurements for one time step of several of our examples, broken down by (1) generating surface and balance particles, (2) transferring MPM particle quantities
to the background grid and implicit grid update, and (3) transferring background grid quantities to MPM particles, merging surface, balance and MPM particles, and updating MPM particle states. Note that the merging time does not exceed $3.5\%$ of (3).  All times are in milliseconds.}
\resizebox{\columnwidth}{!}{%
\begin{tabular}{@{}lllllll@{}}
\toprule
Example & \# Cells & \# Int. Part. & \# Surf. Part. & Sampling & Part.$\rightarrow$Grid & Grid$\rightarrow$Part. \\ \midrule
Droplet Impact ($k^{\sigma} = 5$) & 2M & 794K & 100K & 2224 & 7705 & 2360 \\
Droplets on Ramps ($k^\sigma_{SL}/k^\sigma_{LG} = 0.05$) & 1.5M & 70K & 100K & 258 & 1147 & 287 \\
Contact Angles ($k^\sigma_{SL}/k^\sigma_{LG} = 0$) & 256K & 230K & 250K & 492 & 3715 & 571 \\
Soap Droplet in Water & 1M & 4M & 200K & 2166 & 32099 & 3205 \\
Wine glass & 2M & 1.6M & 500K & 1549 & 11268 & 1172 \\
Candle ($k^{\sigma} = 0.1$) & 2M & 618K & 50K & 1420 & 27500 & 1662 \\
Candle Letters & 256K & 3.1M & 100K & 4601 & 170048 & 2739 \\
Droplet with Marangoni Effect & 4.1M & 235K & 200K & 29991 & 8812 & 3244 \\\bottomrule
\end{tabular}
}
\label{tbl:perf}
\end{table}

\section{Discussion and Future Work}
Our method allows for simulation of surface tension energies with spatial gradients, including those driven by variation in temperature. Our MPM approach to the problem resolves many interesting characteristic phenomena associated with these variations. However, while we provide for perfect conservation of linear and angular momentum, our approach to the thermal transfers is not perfectly conservative. Developing a thermally conservative transfer strategy is interesting future work. Also, although we simulate tears of wine on the walls of a glass, we did not simulate the effect of alcohol evaporation on the surface energy variation. Adding in a mixture model as in \cite{ding:2019:thermomechanical} would be interesting future work. Lastly, although our approach was designed for MPM, SPH is more commonly used for simulation of liquids. However, SPH and MPM have many similarities as recently shown by the work of Gissler et al. \cite{gissler:2020:implicit} and it would be interesting future work to generalize our approach to SPH.

% DO NOT INCLUDE ACKNOWLEDGMENTS IN AN ANONYMOUS SUBMISSION TO SIGGRAPH 2019
%\begin{acks}
%
%The authors would like to thank Dr. Maura Turolla of Telecom
%Italia for providing specifications about the application scenario.
%
%The work is supported by the \grantsponsor{GS501100001809}{National
%  Natural Science Foundation of
%  China}{http://dx.doi.org/10.13039/501100001809} under Grant
%No.:~\grantnum{GS501100001809}{61273304\_a}
%and~\grantnum[http://www.nnsf.cn/youngscientists]{GS501100001809}{Young
%  Scientists' Support Program}.
%
%
%\end{acks}

% Bibliography
\bibliographystyle{ACM-Reference-Format}
\bibliography{../references}

% Appendix
%\appendix
%\section{My Appendix}

%TODO

\end{document}